\documentclass[a4paper,11pt,feynmf]{article}
\usepackage{amssymb}
\usepackage{amsmath,amssymb,mathrsfs}
\usepackage{graphicx}
\usepackage{subcaption}
\usepackage[numbers,sort&compress]{natbib}
\usepackage{ifpdf}
\usepackage{xcolor}
\usepackage{float}
\usepackage[top=30mm, right=30mm, bottom=20mm, left=30mm]{geometry}
\newcommand*{\rom}[1]{\expandafter\@slowromancap\romannumeral #1@}
\begin{document}

\vskip .3in {\textbf{\Large{Transverse expansion of  (1 + 2) dimensional  magneto-hydrodynamics flow with  longitudinal boost  invariance }}}
\medskip

\vskip .3in \centerline{A.~Emamian, A. F.~Kord, A.~Ghaani, B.~Azadegan }
\bigskip
{\small{\it { Department of Physics, Hakim Sabzevari University (HSU), P.O.Box 397, Sabzevar, Iran.}}}

\small{\it\center{ E-mail: $afarzaneh@hsu.ac.ir$}} \vskip .3in

\begin{abstract}
In the present work, we investigate the effects of  magnetic field  on expanding hot and dense nuclear matter as an ideal fluid. We consider QGP, on the particular case of a (1 + 2) dimensional longitudinally boost-invariant fluid expansion, in the background of an inhomogeneous magnetic field that is generated by external sources. We assume the magnetic  field points in the direction perpendicular to the reaction plane, follows the power-law decay in proper time, and has two components on the transverse plane.
To simplify the calculation, we suppose the investigated fluid has azimuthal symmetry, and magneto-hydrodynamic equations are described in a polar coordinate system on the transverse plane of reaction. Our results depict the space-time evolution of the transverse expansion of the fluid in the presence of an inhomogeneous external magnetic field.  Ultimately, we utilize transverse velocity and correction of energy density to estimate the transverse momentum spectrum of final particles that emerge from heavy-ion collisions based on  experimental data.
\end{abstract}

\section{Introduction}
According to the experimental data from relativistic Heavy Ion Collisions (HICs) at RHIC and LHC, a unique form of hot and dense nuclear matter is in the initial stage of collisions created, which is ordinarily known as Quark-Gluon Plasma (QGP). It is discussed that QGP behaves more like a strongly coupled, nearly perfect fluid. The QGP phase description in terms of relativistic hydrodynamics has been fairly successful in the heavy-ion collision experiments \cite{a1,a2,a3,a4,a5,a6}.

One of the interesting phenomena in the heavy-ion collision is generating a strong magnetic field in the early stages of heavy-ion collision due to the spectator protons' relativistic motion. It is argued the strength of this magnetic field can be reached to the order of $eB\sim m_{\pi}^2\sim 10^{18}-10^{19} G$ in the initial stages of the collision \cite{a7,a8,a9}.

The significance of EM fields on the QGP medium has been investigated for some events. For example, if the number of left- and right-handed quarks is unbalanced, a net charge current will be induced by the magnetic field. This phenomenon is called chiral magnetic effect (CME) \cite{a8,a10}. Besides, chiral magnetic wave (CMW) is another phenomenon that one could describe in terms of a chiral current induced by the magnetic field \cite{a11}. These effects could explain breaking symmetry in the angular distribution of final charged particles, and also the difference between the elliptic flow of $\pi^{+}$ and the elliptic flow of $\pi^{-}$ \cite{a12}.
Some phenomena may be explained in terms of an electric field in the early stage of QGP, such as the chiral electric separation effect and chiral Hall-separation effect. One can find more references and information in \cite{a13,a14,a15,a16}.

To study the phenomena mentioned above in the relativistic hydrodynamic framework, one needs to solve the Maxwell equations and the conservation equations for the energy-momentum tensor. The model is called the relativistic magneto-hydrodynamic (RMHD). There is an open question about the effect of the magnetic field on the evaluation of QGP. It is known the magnetic field generated in heavy-ion collisions decays very fast in the vacuum \cite{a8}, but its decay becomes more slowly in the presence of an electrically conducting media \cite{a17,a18,a19,a20}. Therefore, there is still a challenge of how magnetic field may have significant consequences on the quark-gluon matter dynamics. Simplified studies have been done about a magnetic field's possible effect on QGP evolution \cite{a18,a19}. Besides, more numerical advance methods have been used in the framework of the relativistic magneto-hydrodynamic (RMHD) or a (1+3) dimensional partonic cascade BAMPS (Boltzmann Approach to Multi-Parton Scatterings) \cite{a21,a22,a23,a24}.

Moreover, several works have been published in a parallel analytical approach, with longitudinal boost invariance and an external transverse magnetic field \cite{a25,a26,a27,a28}. Solutions have been found a (1+1) dimensional Bjorken flow within ideal transverse RMHD. Furthermore, in ref \cite{a26}, the magnetization property has been added and studied. The (1+2) dimensional expansion of  a perfect fluid  in the presence of an external magnetic   field  has been examined in  \cite{a29}, where the external magnetic  field has been  pointed  in the  transverse  Cartesian coordinate $y$ direction.  Besides,  all hydrodynamic quantities ($P, \epsilon, T, B^2$)  have been dependent on proper time ($\tau$) and  Cartesian coordinates ($x,y$). The model has been called reduced MHD.

In an earlier paper \cite{a28}, we have considered an ideal fluid in the presence of a weak external inhomogeneous magnetic field pointing along the $\phi$ direction. Besides, we have discussed the expansion in a cylindrical coordinate system. Because of the particular choice of the magnetic field ($\phi$ direction), we have only considered central collisions. The decay of the magnetic field has been assumed as $\tau^{n/2}$ with $\tau$ being proper time and $n$ being an arbitrary negative number. The analytical solutions have been only obtained for $n=-1$ and $n=-\frac{4}{3}$. In recent work, we shall modify the previous model and consider a weak external inhomogeneous magnetic field pointing along the arbitrary direction in the transverse plane; therefore, the magnetic field has two components $b_{\phi}$ and $b_r$ in a cylindrical coordinate system. Thus, we investigate the (1+2) dimensional expansion of an ideal perfectly conducting fluid in the hydrodynamic framework in the presence of the magnetic field. For simplicity, we ignore magnetization and dissipative effects. It is known for a perfectly conducting fluid the evolution of energy density, and magnetic field are decoupled because of frozen-flux theorem \cite{a26,a30}. However, we follow the scenarios in refs \cite{a25,a26,a27}, and consider a conducting fluid in the presence of an external magnetic field with power decay low as $\tau^{n/2}$ and $n$ being a negative number of the fluid. In the heavy-ion collisions, we mostly consider outlines in which the magnetic field decays more rapidly than in the ideal-MHD case, and energy density decays more slowly than in the ideal-MHD case. The ideal-MHD point corresponds to $n=-2$.

The paper is organized as follows. In section 2, we illustrate the ideal relativistic magneto-hydrodynamic framework  in the case of a plasma. Next, we display our perturbative approach and  obtain  analytical solutions. The general results  are discussed in  section 3. Then, we calculate the transverse momentum spectrum and compare it with experimental data obtained in RHIC. The last section relates the conclusions and subsequent issues.

 \section{Ideal relativistic magneto-hydrodynamic in the transverse expansion}
We present a brief preview of our formalism to describe the interaction of matter and electromagnetic fields in plasma. We assume there is only an external magnetic field interacted with QGP matter. The external magnetic field is generated by the charged spectators. It is known external fields may induce internal electromagnetic fields dictated by Maxwell’s equations. To give a complete description of matter, one should solve both Maxwell’s equations and the conservation equations together. However, in the present work, we ignore the induced internal electromagnetic fields and only investigate the effects of the external magnetic field on QGP matter. Besides, we assume an arbitrary function for the magnetic field because external sources create it, and we only solve the conservation equations.

We discuss the case of an ideal non-resistive plasma with massless particles. Besides, the set of equations can be closed by incorporating a thermodynamic EoS assumed the pressure is merely proportional to the energy density as $P = \kappa\epsilon$ where $\kappa$ is constant. Finally, we present the the energy-momentum conservation equations for an ideal fluid in the presence of an external magnetic field in the RMHD framework \cite{a31,a32}.

The covariant form of energy-momentum conservation equations is given by:
\begin{eqnarray}\label{Con}
d_\mu(T_{F}^{\mu\nu}+T_{EM}^{\mu\nu})=0
\end{eqnarray}
where
\begin{eqnarray}
T_{F}^{\mu\nu}&=&(\epsilon+P)u^\mu u^\nu+Pg^{\mu\nu}\\
T_{EM}^{\mu\nu}&=&b^2 u^\mu u^\nu+\frac{1}{2}b^2 g^{\mu\nu}-b^\mu b^\nu
\end{eqnarray}
Here $\epsilon$ and $P$ are the energy density and pressure of the fluid, respectively, and  $b^{\mu}$ is  the magnetic field four-vector with modulus $b^{\mu}b_{\mu}=b^2$. The metric tensor in a flat spacetime is  $g_{\mu \nu}=diag \lbrace-,+,+,+\rbrace$, and  so that  $u_{\mu}=\gamma(1, \vec v)$ is the  four-velocity of fluid satisfying $u^\mu u_\mu=-1$.

By introducing the orthogonal projector to the fluid four-velocity as $\Delta^{\mu\nu}=g^{\mu\nu}+u^\mu u^\nu$, one can rewrite the conservation equation \label{Con} alongside the parallel and perpendicular directions to $u_{\nu}$. They are:
\begin{eqnarray}\label{en}
u_\nu d_\mu(T_{F}^{\mu\nu}+T_{EM}^{\mu\nu})=0 \Longrightarrow D(\epsilon+\frac{b^2}{2} )+(\epsilon+P+b^2)\Theta+u_\nu b^\mu(d_\mu b^\nu)=0\nonumber\\
\end{eqnarray}
and
\begin{eqnarray}\label{Eu}
\Delta_{\alpha \nu} d_\mu(T_{F}^{\mu\nu}+T_{EM}^{\mu\nu})=0\Longrightarrow(\epsilon+P+b^2)Du_{\alpha}+\nabla_{\alpha}(P+\frac{b^2}{2})-d_\mu (b^\mu b_{\alpha})-u_{\alpha}u_\nu d_\mu( b^\mu b^\nu)=0\nonumber\\
\end{eqnarray}
where:
\begin{eqnarray}
D=u^\mu d_\mu,\ \ \  \Theta=d_\mu u^\mu,\ \ \  \nabla_\alpha=\Delta_{\alpha\nu}d^\nu
\end{eqnarray}
In relativistic heavy-ion collisions, the nuclei are accelerated toward each other in the center-of-mass frame with velocities  near the light speed. However, all the motion is longitudinal before the collision, and it may still be assumed to remain in the longitudinal direction. Moreover, the fluid is assumed with a finite transverse size and expands both in the longitudinal and radial direction. Thus the four-velocity for an ideal fluid  following the Bjorken expansion along the $z$ direction and moving in the transverse plane only in the $r$ direction in a cylindrical coordinate becomes:
\begin{eqnarray}
u_\mu=\frac{1}{\sqrt{1-u_z^2-u_r^2}}(1, u_r, 0, u_z)
\end{eqnarray}
where $u_r$ and $u_z$ are the radial and longitudinal components, and  $u_\phi$ is zero. Besides, the boost invariance of the Bjorken expansion allows us to restrict the discussion to the $z =0$ plane where symmetry reasons impose $u_z$ =0.  As we know, it is more advantageous to work in Milne coordinates, $x_{\mu}=(\tau, r, \phi, \eta)$, such that:
\begin{eqnarray}
\tau=\sqrt{t^2-z^2},\quad \eta=\frac{1}{2}\ln\frac{t+z}{t-z},\quad \phi=\tan^{-1}(y/x),\quad r^2=x^2+y^2
\end{eqnarray}
and the metric for these coordinates is parameterized as $g_{\mu\nu}=diag\lbrace-1, 1,r^2, \tau^2\rbrace$.
The covariant derivatives in Eqs.~(\ref{en}) and (\ref{Eu}) are defined as:
\begin{eqnarray}
&&d_\mu A^\nu=\partial_\mu A^\nu+\Gamma^\nu_{\mu\rho}A^\rho \nonumber\\
&&d_\mu(A^\nu A^\alpha)=\partial_\mu(A^\nu A^\alpha) +\Gamma^\nu_{\mu\sigma}A^\sigma A^\alpha + \Gamma^\alpha_{\mu\sigma}A^\nu A^\sigma
\label{derivative}
\end{eqnarray}
where the only non-zero Cristoffel symbols according to the following definition:
\begin{equation}
\Gamma^i_{jk}=\frac{1}{2}g^{im}\Big(\frac{\partial g_{mj}}{\partial x^k}+\frac{\partial g_{mk}}
{\partial x^j}-\frac{\partial g_{jk}}{\partial x^m}\Big).
\end{equation}
are $\Gamma^\tau_{\eta\eta}=\tau,\ \Gamma^{r}_{\phi\phi}=-r, \ \Gamma^\phi_{r\phi}=\frac{1}{r},\
\Gamma^\eta_{\tau\eta}=\frac{1}{\tau}$.

In order to simplify our calculation, we apply the longitudinal boost invariance and rotational symmetry around the beamline. The assumptions of boost invariance and rotational symmetry around the beamline imply that none of the quantities depends on $\eta$ and $\phi$ coordinates. Consequently, we expect that all the quantities of interest solely rely on the transverse radial $r$ and $\tau$ coordinate in the $(\tau,r,\phi,\eta)$ coordinate system. The assumption of boost invariance is approximate correct near mid rapidity \cite{a32}. Moreover, it is known collisions with significant impact parameters break the rotational symmetry. In this case, to describe systems without rotational symmetry, one needs to include three additional quantities: the shear and bulk viscosities and the thermal conductivity, respectively~\cite{a34}. However, a general calculation, including the viscosities, is out of the scope of this work.

To improve the previous work of two of the authors \cite{a28}, due to non-central collisions  and asymmetric charge distribution, we suppose that the magnetic field is perpendicular to the reaction plane pointing along the $r$ and $\phi$ directions.  However, we assume the magnitude of the magnetic field is  independent of  $\phi$, but it is shown each component of the magnetic field can be dependent on the $\phi$ coordinate.

We also ignore the Maxwell equations and assume the external sources govern the dynamics of the magnetic field. Then, we solve the conservation equations perturbatively and analytically. It is known the magnitude of the magnetic field drops rapidly concerning time because fast velocities of the charged spectators (as sources for the external magnetic field) fly away along the beam direction \cite{a8,a17}. Thus, the energy density of the magnetic field could be suppressed by the energy density of the fluid at some initial time after collisions \cite{a27}. Therefore, we follow the perturbative scenario which has been introduced in \cite{a27}, and consider $\frac{b^2}{\epsilon}\ll 1$. Hence, one can ignore Maxwell equations, and external sources dictate the dynamics of the magnetic field. Furthermore, as been discussed by authors in \cite{a27}, one could ignore the nonlinear effects in $b^2$ at a perturbative treatment despite that they may be necessary in very early times.

We shall investigate  perturbative solutions of the conservation equations in the presence  of an external  weak  inhomogeneous magnetic field pointing along  the $r$ and $\phi$ directions  in an inviscid fluid with a high enough electrical conductivity. The initial setup of our study is given by:
\begin{eqnarray}
\label{in}
&& u_\mu=(1, \lambda^2 u_r, 0, 0) , \qquad  b_\mu=(0, \lambda b_r, \lambda r b_\phi, 0), \qquad \epsilon=\epsilon_0(\tau)+\lambda^2\epsilon_1(\tau, r)\nonumber\\
\end{eqnarray}
Here  $\epsilon_0(\tau)=\frac{\epsilon_c}{\tau^{4/3}}$, and  $\tau$  is rescaled by an initial time $\tau_0$. $\epsilon_c$ is the energy density at proper time $\tau_0$. Also, $\lambda$ is an expansion parameter, which will be set to unity at the end of calculations. Besides, in our calculations, we implicitly  rescale $r$ by $\tau_0$.
By substituting the above condition in the energy conservation (\ref{en}) and Euler equation (\ref{Eu}), we have obtained  three coupled differential equations. Up to order $O(\lambda^2)$, we have:
\begin{eqnarray}
\label{E1}
 &&\partial_{\tau}\epsilon_1 -\frac{4 \epsilon_c}{3\tau^{4/3}}\left(\frac{u_r}{r}+\frac{\partial
u_r}{\partial r}\right)+\frac{4\epsilon_1}{3\tau}+\frac{b^2}{\tau}+\frac{1}{2}\partial_\tau b^2=0\\
\label{E3}
&&\partial_r(b_r b_\phi) + \frac{2b_r b_\phi}{r}+\frac{2b_\phi}{r}\partial_\phi b_\phi =0\\
\label{E2}
&&\partial_r\epsilon_1-\frac{4\epsilon_c}{\tau^{4/3}}\partial_\tau u_r+\frac{4\epsilon_c}
{3\tau^{7/3}}u_r-\frac{3}{r}(b_r^2-b_\phi^2)+\frac{3}{2}\partial_r b^2-3\partial_r b_r^2
-\frac{3}{r}\partial_\phi(b_\phi b_r)=0\nonumber\\
\end{eqnarray}
Assuming rotational symmetry around the beamline:
\begin{eqnarray}
\label{EE1}
\partial_\phi b^2(\tau ,r)=0
\end{eqnarray}
and using the method  of the  separation of variables, we have solved Eqs.~(\ref{E3}) and~(\ref{EE1}). Then,  $b_r, b_\phi$, and $b^2$ are quickly achieved as follow:
\begin{eqnarray}
\label{EE2}
 b_r(\tau ,r,\phi)&=&\sum_{l=-2}^{\infty}C_l(\tau)r^{\frac{l}{2}}[A_lcos(\frac{(l+2)\phi}{2})+B_lsin(\frac{(l+2)\phi}{2})] \\
 \label{EE3}
 b_\phi(\tau ,r,\phi) &=&\sum_{l=-2}^{\infty}C_l(\tau)r^{\frac{l}{2}}[B_lcos(\frac{(l+2)\phi}{2})-A_lsin(\frac{(l+2)\phi}{2})]\\
 \label{EE4}
b^2(\tau ,r)&=&\sum_{l=-2}^{\infty}C_l(\tau)r^{l}\{[A_lcos(\frac{(l+2)\phi}{2})+B_lsin(\frac{(l+2)\phi}{2})]^2 \nonumber\\ && + [B_lcos(\frac{(l+2)\phi}{2})-A_lsin(\frac{(l+2)\phi}{2})]^2\}\nonumber\\ &=&\sum_{l=-2}^{\infty}C^{\prime}_l(\tau)r^l=f(r,\tau)
\end{eqnarray}
where $l  \geq -2$ are real numbers, and $C_l,A_l,B_l$ and $C^{\prime}_l$ are expansion constants. These expansion coefficients must be obtained according to the physical condition of the problem. According to the existing conditions in the case of an  external  magnetic field, all coefficients with index $l < 0$ should be zero. Then, we substitute $b_r, b_\phi$, and $b^2$ from above results in Eqs.~(\ref{E1}) and (\ref{E2}), and  obtain:
\begin{eqnarray}
\label{E6}
&&\partial_{\tau}\epsilon_1 -\frac{1}{3\tau r}\frac{\partial}{\partial r}\left(  \frac{4\epsilon_cru_r}{\tau^{1/3}}\right)+\frac{4\epsilon_1}{3\tau}+\frac{b^2}{\tau}+\frac{1}{2}\partial_\tau b^2=0\\
\label{E7}
&&\partial_r\epsilon_1-\frac{1}{\tau r}\frac{\partial}{\partial \tau}\left(  \frac{4\epsilon_cru_r}{\tau^{1/3}}\right)=0
\end{eqnarray}
Interestingly, using results in Eqs.~(\ref{EE2})-(\ref{EE4}) we could simplify  Eqs.~(\ref{E1}), (\ref{E2}) as  Eqs.~(\ref{E6}) and (\ref{E7}).
One could combine the above equations, and obtains  a partial differential equation depending on $u_r(\tau ,r)$ and $b^2(\tau ,r)$:
\begin{eqnarray}
\label{E8}
&&\frac{3 r^2 \tau ^{7/3}}{4 \epsilon_c}\partial_r b^2(\tau ,r)+\frac {3 r^2 \tau ^{10/3}}{8 \epsilon_c}\partial_r\partial_\tau b^2(\tau ,r)-r^2 \tau^2 \partial_r^2 u_r(\tau ,r)+3 r^2 \tau ^2 \partial_\tau^2 u_r(\tau ,r)\nonumber\\&&-r^2 \tau  \partial_\tau u_r(\tau ,r)+r^2 u_r(\tau ,r)-r \tau ^2 \partial_r u_r(\tau ,r)+\tau ^2 u_r(\tau ,r)=0.
\end{eqnarray}
This equation describes the dynamical evolution of the transverse velocity of the fluid in the presence of an external magnetic field according to our initial setup Eq.~(\ref{in}), and its solution directs us to explain the fluid velocity $u_r(\tau ,r)$ in terms of the cylindrical radial coordinate.

\subsection{Analytical solution}

To solve Eq.~(\ref{E8}), we first assume $b^2(\tau ,r)=0$ condition and obtain the general solution for the homogeneous partial differential equation by the manner of separation of variables. The general solution is given by:
\begin{eqnarray}
\label{E9}
u_r^h(\tau, r)&=&\sum _m \Big(A_1^m J_1(m r)+A_2^m Y_1(m r)\Big)\nonumber\\&& \times\Big(\tau ^{2/3} A_3^m J_{\frac{1}{3}}({m \tau }/{\sqrt{3}})+\tau ^{2/3} A_4^m Y_{\frac{1}{3}}({m \tau }/{\sqrt{3}})\Big)
\end{eqnarray}
where $m$ can be real or imaginary numbers and $A_{1,2,3,4}^m$ are integration constants. To find the solution for $b^2(\tau ,r)\neq0$, we make some assumptions and   converting   the partial differential equation into the summation of the solution of ordinary differential equations. We consider $b^2(\tau, r)$ into a series based on the $r$-dependence part of the solution and assume the time dependence of $b^2(\tau, r)$ behavior as $\tau^n$ with $n<0$ which nearly represents the decay of magnetic field in heavy-ion collisions \cite{a25,a26,a27}:
\begin{eqnarray}
\label{E10}
b^2(\tau, r)=\sum_m \tau^n B_m^2\ f(m r)
\end{eqnarray}
here $B_m^2$ are constants. Also, the following ansatz for radial velocity is considered:
\begin{eqnarray}
\label{E11}
u_r(\tau, r)=\sum_m\Big(a_m(\tau)J_1(m r)+b_m(\tau)Y_1(m r)\Big)
\end{eqnarray}
It is known  that initial condition $u_r(\tau =0, r)=0$ leads to $b_m(\tau)=0$. Substituting the Eqs.~(\ref{E10}) and (\ref{E11}) into Eq.~(\ref{E8}), we obtain:
\begin{eqnarray}
\label{E12}
 J_1(m r) \Big((m^2 \tau ^2+1) a_m(\tau )+3\tau^2 a_m''(\tau )-\tau a_m'(\tau )\Big)+\frac{3 m (n+2)  B_m^2 \tau ^{n+\frac{7}{3}}}{8 \epsilon_c} f'(m r)=0
\end{eqnarray}
We can reach the following ordinary differential equation for the function $f(m r)$ by separation of variables:
\begin{eqnarray}
\label{E13}
f'(m r)=-J_1(m r)
\end{eqnarray}
its general solution becomes:
\begin{eqnarray}
\label{E14}
f(m r)= c_1+J_0(m r)
\end{eqnarray}
For simplicity, we impose $c_1=0$, but in a realistic situation, one should obtain it from physical conditions on the magnetic field. However, the magnetic field should be prominent at $r=0$. Using Eqs.~(\ref{E10}) and (\ref{E14}), we can expand the $b^2(\tau, r)$ in terms of zero-order Bessel functions as following:
\begin{eqnarray}
\label{E15}
b^2(\tau, r)=\sum_{m} \ \tau^{n}\ B_m^2  J_0(\alpha_{0 m} \frac{r}{a})
\end{eqnarray}
where the coefficients $B_m^2$ are given by:
\begin{eqnarray}
\label{E16}
B_m^2=\frac{2}{a^2[J_1(\alpha_{0 m})]^2}\int_0^a \ r J_0(\alpha_{0 m}\frac{r}{a}) \ b^2(\tau, r) \ dr
\end{eqnarray}
here $\alpha_{0 m}$ is the $m$th zero of $J_0$.

Besides, the transverse velocity is given by:
\begin{eqnarray}
\label{E17}
u_r(\tau, r)=\sum_m a_m(\tau) J_1( m r)
\end{eqnarray}
Thus, the coefficients $a_m(\tau)$  is obtained by solving  the following ordinary differential equation:
\begin{eqnarray}
\label{E18}
3\tau^2 a_m''(\tau )-\tau a_m'(\tau )+(m^2 \tau ^2+1) a_m(\tau )-\frac{3 m (n+2)  B_m^2 \tau ^{n+\frac{7}{3}}}{8 \epsilon_c}=0
\end{eqnarray}
and the analytical solution is:
\begin{eqnarray}
\label{E19}
a_m(\tau)&=&\tau ^{2/3}\Big(c_1^m J_{\frac{1}{3}}(\frac{m \tau }{\sqrt{3}})+c_2^m Y_{\frac{1}{3}}(\frac{m \tau }{\sqrt{3}})\Big)+\frac{3 m \tau ^{n+\frac{7}{3}}B_m^2}{16 \epsilon_c (3 n+4)}\nonumber\\&&
 \Big(3 (n+2) \, _0F_1[\frac{4}{3},-\frac{1}{12} m^2 \tau ^2] \, _pF_q[\{\frac{n}{2}+\frac{2}{3}\},\{\frac{2}{3},\frac{n}{2}+\frac{5}{3}\},-\frac{1}{12} m^2 \tau ^2]-(3 n+4)\nonumber\\&& \, _0F_1[\frac{2}{3},-\frac{1}{12} m^2 \tau ^2] \, _pF_q[\{\frac{n}{2}+1\},\{\frac{4}{3},\frac{n}{2}+2\},-\frac{1}{12} m^2 \tau ^2]\Big)
\end{eqnarray}

The transverse velocity is completely determined if constants $c_1^m$ and $c_2^m$ are fixed. We can find $c_1^m$ and $c_2^m$ by initial condition at $\tau\rightarrow\infty$. Since $b^2(\infty, r)\rightarrow0$, we expect $u_r(\infty, r)\rightarrow0$. By making late-time expansion of $u_r(\tau, r)$, one finds that this takes the asymptotic form as $f(\tau)\tau^{1/6}$ where $f(\tau)$ is an oscillatory function. To prevent divergent transverse velocity, one has to choose the coefficient of $\tau^{1/6}$ equal to zero. The solutions satisfying the initial condition at $\tau\rightarrow\infty$ are written as follows:
\begin{eqnarray}
\label{E20}
&&c_1^m = \frac{\pi ^2 2^{n-\frac{7}{3}} 3^{\frac{n}{2}+\frac{1}{3}} B_m^2 m^{-n-\frac{2}{3}} \Big(\csc (\frac{\pi  n}{2})+2 \sec (\frac{1}{6} (3 \pi  n+\pi ))\Big)}{\epsilon _c \Gamma (-\frac{n}{2}-1) \Gamma (\frac{1}{3}-\frac{n}{2})}
 \nonumber\\
&&c_2^m = -\frac{\pi  2^{n-\frac{7}{3}} 3^{\frac{n}{2}+\frac{5}{6}} B_m^2 m^{-n-\frac{2}{3}} \Gamma (\frac{n}{2}+2)}{\epsilon _c \Gamma (\frac{1}{3}-\frac{n}{2})}
\end{eqnarray}

After finding $u_r(\tau, r)$ we could obtain the modified energy density from Eq.~(\ref{E7}). It is given by:
\begin{eqnarray}
\label{E21}
\epsilon_1 (\tau,r)= \sum_m h^m(\tau)-\sum_m\frac{4 \epsilon _c (1-J_0(m r)) (a_m(\tau )-3 \tau  a_m'(\tau ))}{3 m \tau ^{7/3}}
\end{eqnarray}
In the above equation, $h^m(\tau)$ is the constant of integration which is determined  by solving  Eq.~(\ref{E6}):
\begin{eqnarray}
\label{E22}
h^m(\tau)=\frac{\int_1^{\tau } (\frac{4}{3} \epsilon _c m a_m (s)-\frac{1}{2} B_m^2 (n+2) s^{n+\frac{1}{3}}) \, ds}{\tau ^{4/3}}
\end{eqnarray}
Finally, replacing $a_m(\tau)$ and its derivative into Eq.~(\ref{E21}) lead us to determine  the modified energy density as:
\begin{eqnarray}
\label{E23}
&&\epsilon_1 (\tau,r)=\sum_m h^m(\tau) - \sum_m\frac{4 \epsilon _c (1-J_0(m r))}{3 \tau ^{7/3} m}
\Big(-\sqrt{3} \tau ^{5/3} m ( c_1^m J_{-\frac{2}{3}}(\frac{m \tau}{\sqrt{3}})+ c_2^m Y_{-\frac{2}{3}}(\frac{m \tau}{\sqrt{3}}))\nonumber\\&&+\frac{1}{64 \epsilon _c} m \tau ^{n+\frac{7}{3}} B_m^2 (-m^2 \tau ^2 \, _0F_1[\frac{5}{3},-\frac{1}{12} \tau ^2 m^2] \, _pF_q[\{\frac{n}{2}+1\},\{\frac{4}{3},\frac{n}{2}+2\},-\frac{1}{12} \tau ^2 m^2]\nonumber\\&&-\frac{8 (n+2) \, _0F_1[\frac{1}{3},-\frac{1}{12} \tau ^2 m^2] \, _pF_q[\{\frac{n}{2}+\frac{2}{3}\},\{\frac{2}{3},\frac{n}{2}+\frac{5}{3}\},-\frac{1}{12} \tau ^2 m^2]}{3 n+4})\Big)
\end{eqnarray}
Note that the above solutions of  $u_r(\tau,r)$ and $\epsilon_1(\tau,r)$ are invalid for $n =-4/3$, which can be observed from the divergence in the inhomogeneous solution Eq.~(\ref{E17}) and (\ref{E23}). For this particular case, the solutions are shown in the following expressions:
\begin{eqnarray}
\label{E24}
&&u(\tau,r)=\sum_m \frac{\pi  m^{2/3} \Gamma (\frac{1}{3}) B_m^2}{8\ 2^{2/3} \epsilon _c}\Big(\frac{1}{3^{4/3}} \tau ^{2/3} J_{\frac{1}{3}}(\frac{m \tau }{\sqrt{3}})-\frac{1}{3^{5/6}} \tau ^{2/3} Y_{\frac{1}{3}}(\frac{m \tau }{\sqrt{3}})\Big)\nonumber\\&&
+\frac{\pi  m \tau B_m^2}{288 \epsilon _c \Gamma (\frac{4}{3})^2 \sqrt[3]{m \tau }}\Big(-2^{2/3} \sqrt[3]{3} \Gamma (\frac{1}{3}) (m \tau )^{2/3} J_{\frac{1}{3}}(\frac{m \tau }{\sqrt{3}}) \, _pF_q[\{\frac{1}{3}\},\{\frac{4}{3},\frac{4}{3}\},-\frac{1}{12} m^2 \tau ^2]\nonumber\\&&+2^{2/3} 3^{5/6} \Gamma (\frac{1}{3}) (m \tau )^{2/3} Y_{\frac{1}{3}}(\frac{m \tau }{\sqrt{3}}) \, _pF_q[\{\frac{1}{3}\},\{\frac{4}{3},\frac{4}{3}\},-\frac{1}{12} m^2 \tau ^2]\nonumber\\&&
-4 \sqrt[3]{2} 3^{2/3} \Gamma (\frac{4}{3})^2 J_{\frac{1}{3}}(\frac{m \tau }{\sqrt{3}}) G_{1,3}^{2,0}(\frac{m^2 \tau ^2}{12}|
\begin{array}{c}
 1 \\
 0,0,\frac{1}{3} \\
\end{array}
))\Big)
\end{eqnarray}
\begin{eqnarray}
\label{E25}
&&\epsilon_1 (\tau,r)=\sum_m h^m(\tau)-\sum _m \frac{4 \epsilon _c (1-J_0(m r))}{3 \tau ^{7/3} m}\Big(-\frac{\pi  m^{5/3} \tau ^{5/3} \Gamma (\frac{1}{3}) B_m^2}{24\ 2^{2/3} \sqrt[3]{3}  \epsilon _c}\nonumber\\&&(\sqrt{3} J_{-\frac{2}{3}}(\frac{m \tau }{\sqrt{3}})-3 Y_{-\frac{2}{3}}(\frac{m \tau }{\sqrt{3}}))+\frac{\pi  (m \tau )^{5/3}}{48\ 2^{2/3} 3^{5/6} \epsilon _c \Gamma (\frac{4}{3})^2}(\sqrt[3]{2} \sqrt[6]{3} (m \tau )^{2/3}\nonumber\\&&\Gamma (\frac{1}{3}) (\sqrt{3} J_{-\frac{2}{3}}(\frac{m \tau }{\sqrt{3}})-3 Y_{-\frac{2}{3}}(\frac{m \tau }{\sqrt{3}})) \, _pF_q[\{\frac{1}{3}\},\{\frac{4}{3},\frac{4}{3}\},-\frac{1}{12} m^2 \tau ^2]\nonumber\\&&+12 \Gamma (\frac{4}{3})^2 J_{-\frac{2}{3}}(\frac{m \tau }{\sqrt{3}}) G_{1,3}^{2,0}(\frac{m^2 \tau ^2}{12}|
\begin{array}{c}
 1 \\
 0,0,\frac{1}{3} \\
\end{array}
))\Big)
\end{eqnarray}
We should consider that in the above equations, $m$ has to be replaced by $\alpha_{0 m}/a$, and one should calculate the integral in Eq.~(\ref{E22}) numerically.

Moreover, for $n=-2$, the evolution of energy density and magnetic field are decoupled. One can see from Eq.~ (\ref{E12}) that the transverse velocity evolution is also decoupled. In this case, the velocity has just the homogeneous solutions, which are assumed to be zero. Thus, the energy correction and the transverse velocity profile is zero.

\section{Results and discussions}
This section will present our results numerically to understand the space-time evolution of quark-gluon plasma in a heavy-ion collision from our perturbative solutions. We would like to present transverse velocity $v_r(\tau, r)=\frac{u^r}{u^\tau}$ and the correction energy density $\epsilon_1$ numerically. It is known the typical magnetic field produced in Au-Au peripheral collisions at $\sqrt{s_{NN}}=200 GeV$ is estimated $ |e B| \thickapprox5 \sim 10 m_{\pi}^{2}$ at $\tau=0$ after a RHIC collision. Assuming the magnitude of the initial energy density of the
medium $\epsilon_c =T^4_0 \thickapprox (300 MeV)^4$ at the initial proper time $\tau_0=0.6$ fm, and also assuming the
initial electromagnetic field reduce to ten times smaller at $\tau=0.6$ fm, one finds $B^2/\epsilon_c \thickapprox 0.17\sim 0.68$ \cite{a27}. Here, we consider $m_\pi\approx150$ MeV and $e^2=4 \pi /137$. In our calculations, we have assumed $B^2/\epsilon_c=0.6$. Note that in our calculations, any change in the ratio $B^2/\epsilon_c$ will only scale the solutions.
\subsection{Numerical solutions}
In order to analyze our perturbative solutions and also obtain some phenomenological insights from them, we consider the following profile of the magnetic field:
\begin{eqnarray}
\label{E26}
b^2(\tau, r)=B_c^2 \tau^{n} e^{- \frac{r^2} {r_0^2}}
\end{eqnarray}
where we assumed the magnetic profile is maximum at $r=0$; besides, the coefficients $r_0$ and $B_c$ are free parameter. So here, $r_0>0$  characterizes the spatial width of the magnetic field.
 This assumed profile allows us to reproduce $b^2$ via a series of Bessel functions as shown in Eq.~(\ref{E15}). The first twenty coefficients $B_m^2$ calculated according to of Eq.~(\ref{E16}) for $r_0=1$ are:
 
$B_c^2$\{0.118491,0.259315,0.364628,0.423534,0.434975,0.40637,0.350437,0.281338,0.211409,0.149238,0.0992222\\,0.0622491,0.0369036,0.0206963,0.0109896
,0.0055289,0.002637,0.00119289,0.000512007,0.000208585\}. 

To reproduce the assumed external magnetic profile from Eq.~(\ref{E26}), taking the first twenty terms of series in the calculation is enough. Note that in all of the following figures, $\tau$, $r$, and $v_r$ are dimensionless, and the dimension of $\epsilon_1$ is $GeV/fm^3$. We remind the reader that $\tau$ and $r$ are rescaled by $\tau_0$.

In the left panel of Fig.~1, we show a comparison between the approximated magnetic field in the Bessel series and the assumed magnetic profile Eq.~(\ref{E26}) at $\tau=1 $ and $r_0=1 $.
In the right panel of Fig.~1, the assumed magnetic profile is plotted as a function of $r$ for different values of $r_0$. As the value of $r_0$  increases,  the external magnetic field is distributed over a broader area of the transverse plane.
In reality, the spatial distribution of the magnetic field depends on the impact parameters of peripheral collisions. In the following calculations, we choose $r_0=1$.

\begin{figure}[h]
		\begin{center}$
	\begin{array}{cc}
     \includegraphics[width=3in]{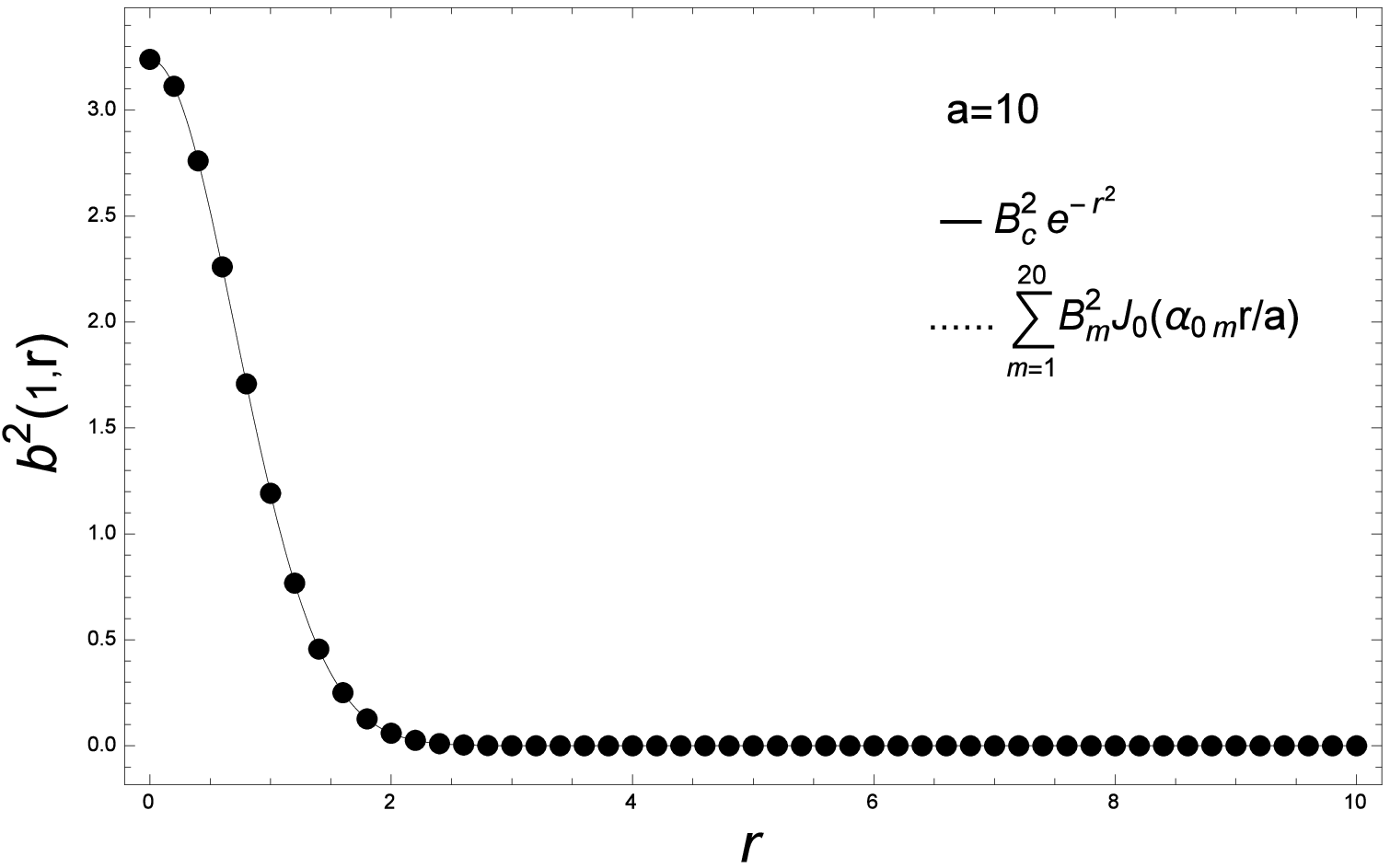}
     \includegraphics[width=3in]{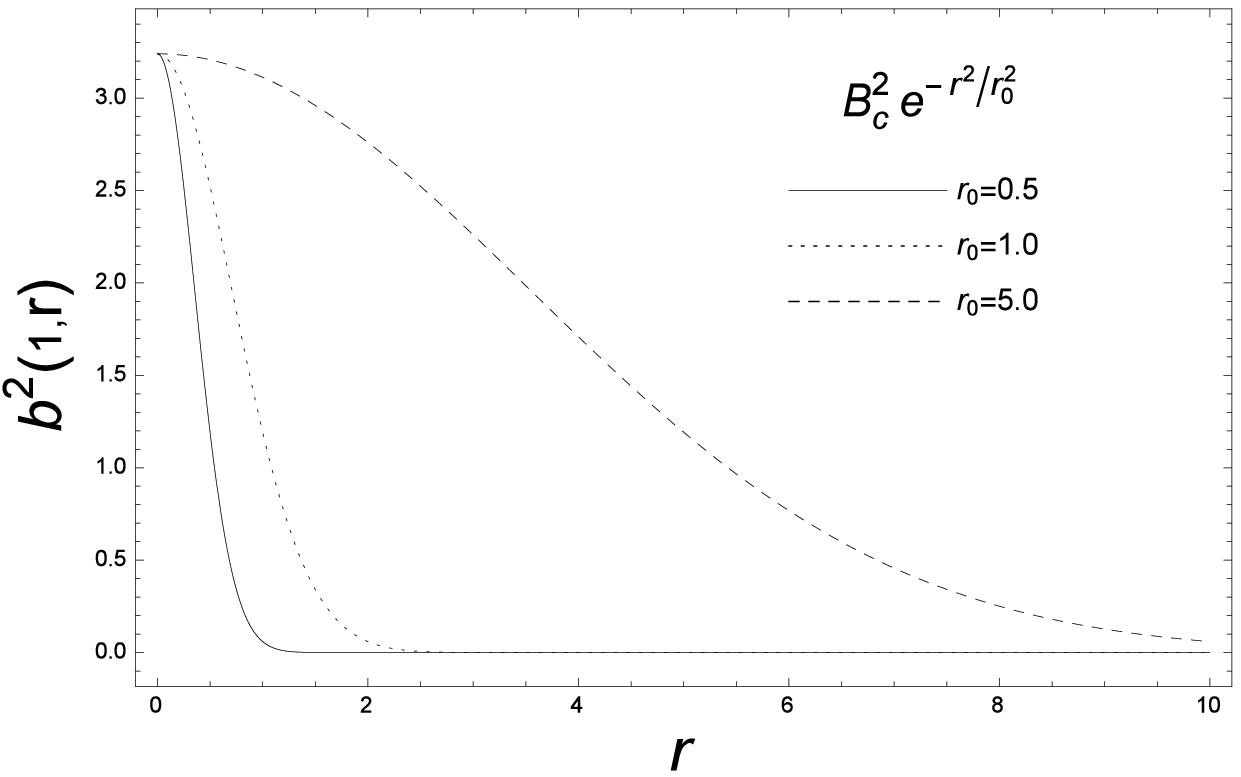}
		\end{array}$
	\end{center}
		\caption  {In the left panel, there is a comparison between the approximated $b^2$ in the Bessel series (dotted curve) and the assumed external magnetic field (solid curve) at $\tau=1$ and $r_0=1$. In the right panel, the assumed external magnetic field is plotted for three different values of $r_0$.}
\end{figure}

In order to demonstrate the effects of the magnetic field on  the fluid velocity  and  the  modified energy density, we plot the transverse fluid velocity and the modified energy density where we have chosen  $B^2_c/\epsilon_c=0.6$. As two examples, we plot the transverse fluid velocity and the modified energy density for two different values of $n=-1,-3$  at either $\tau$ or fixed $r$, respectively. In Figs.~2 and 3, the  transverse flow  $v_r(\tau, r)=\frac{u^r}{u^\tau}=-u_r$ is displayed at  fixed $\tau$ and fixed $r$, respectively. Fig.~2 shows that  the magnitude of the transverse flow      ($\mid v_r(\tau=1,r) \mid$)  increases from $r= 0$, has a maximum at intermediate $r$ and then gradually decreases with $r$.  Fig.~3 shows $\mid v_r(\tau,r=1) \mid$ decreases with respect to proper time due to the decay
 of the magnetic field.

\begin{figure}[h]
	\begin{center}$
\begin{array}{cc}
		\includegraphics[width=4in]{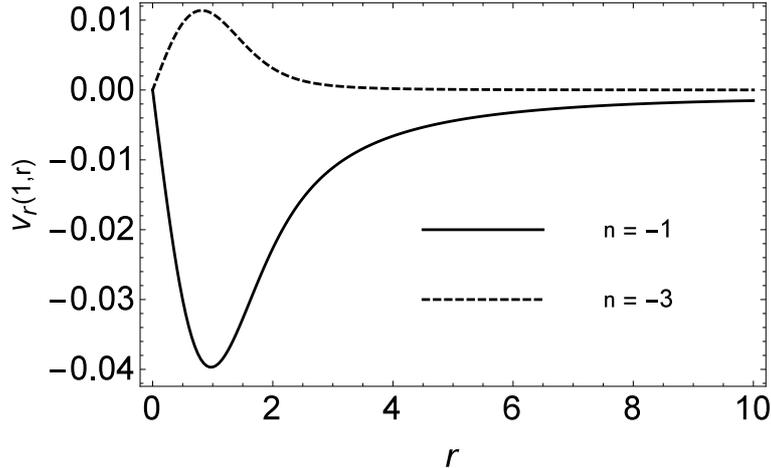}\\
\end{array}$
	\end{center}
	\caption{ The transverse fluid velocity  $v_r(\tau,r)$ in terms of cylindrical radial coordinate $r$  is plotted for two different values of $n=-1,-3$ at $\tau=1$ .}
\end{figure}

\begin{figure}[h]
	\begin{center}$
\begin{array}{cc}
		\includegraphics[width=4in]{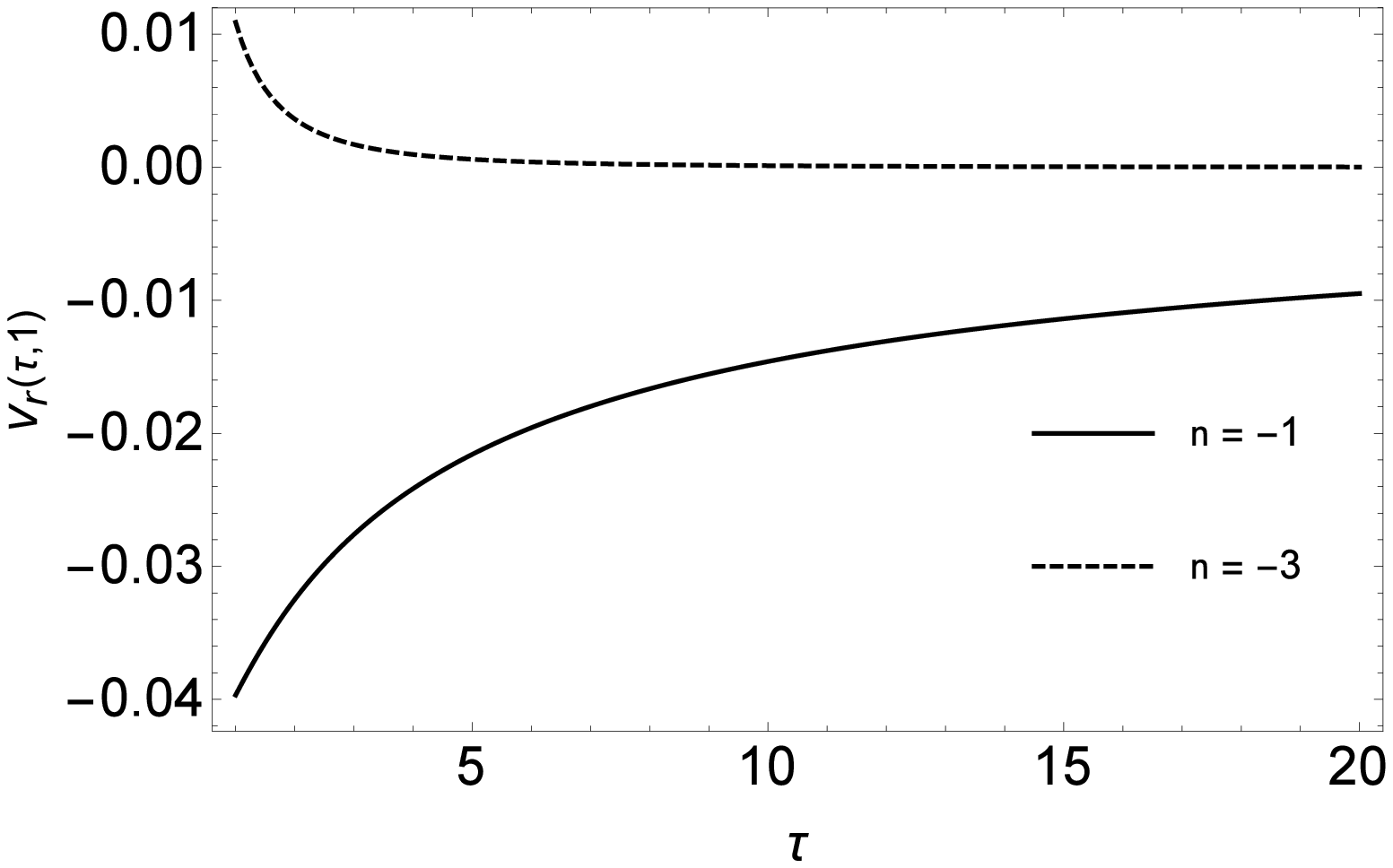} \\
\end{array}$
	\end{center}
\caption{The transverse fluid velocity  $v_r(\tau,r)$ is plotted  as function of  $\tau$  for two different values of $n=-1,-3$ at $r=1$.}
    \end{figure}

in  Figs.~4 and~5 we display $v_r$ in terms of $r$ and $\tau$. Here the horizontal axis and vertical axis correspond to $r$ and $\tau$, respectively. Fig.~4 shows  $v_r$ is pointing inward in the medium for the case  $n=-1$. However, Fig.~5
shows  $v_r$ is pointing outward in the medium  for the case $n=-3$. 
  one could conclude the  external magnetic field causes  
      the transverse compression of fluid for the case $n=-1$, and   the transverse expansion of the fluid  for the case $n=-3$. We should mention that the flow expands longitudinally accordingly to the Bjorken expansion, and it is not shown here. Besides, the transverse flow is more prominent around $r=1$ as shown in Figs.~4 and 5.

\begin{figure}[h]
	\begin{center}$
		\begin{array}{cc}
		\includegraphics[width=4in]{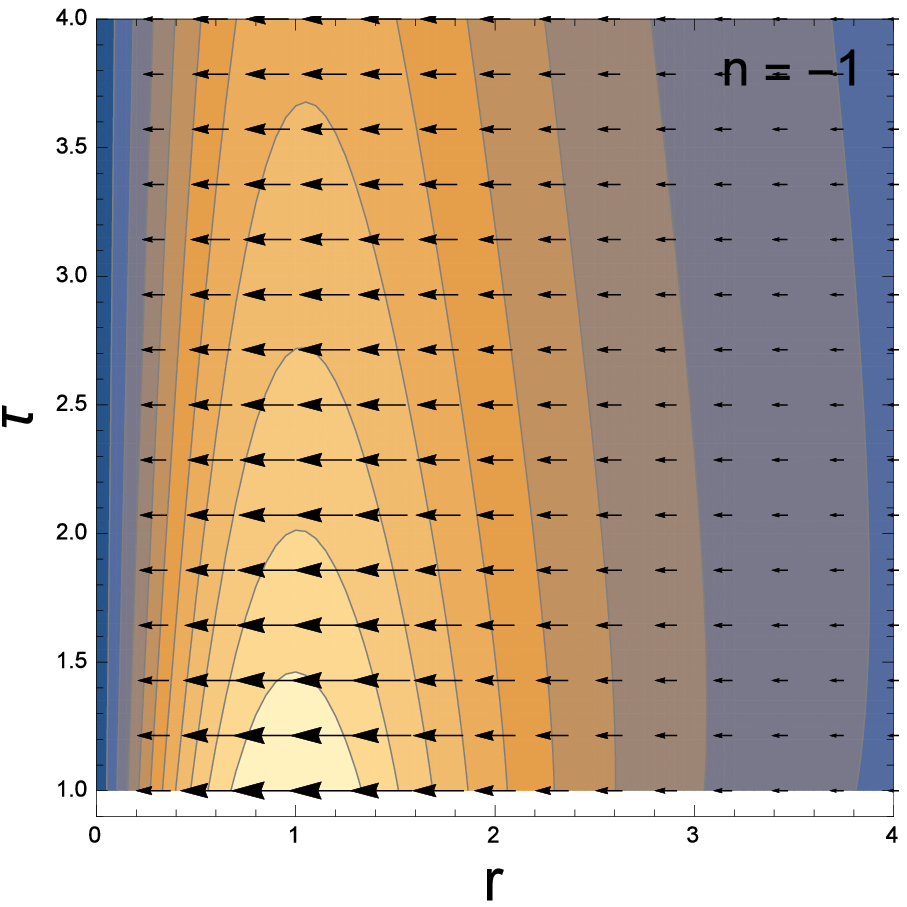}
        \includegraphics[width=1in]{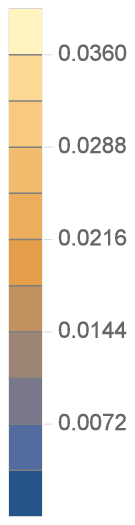}\\
		\end{array}$
	\end{center}
	\caption{ Two-dimensional transverse fluid velocity  $v_r(\tau,r)$ is plotted with $n=-1$. The background colors represent the magnitude of $v_r$.}
\end{figure}

\begin{figure}[h]
	\begin{center}$
		\begin{array}{cc}
		\includegraphics[width=4in]{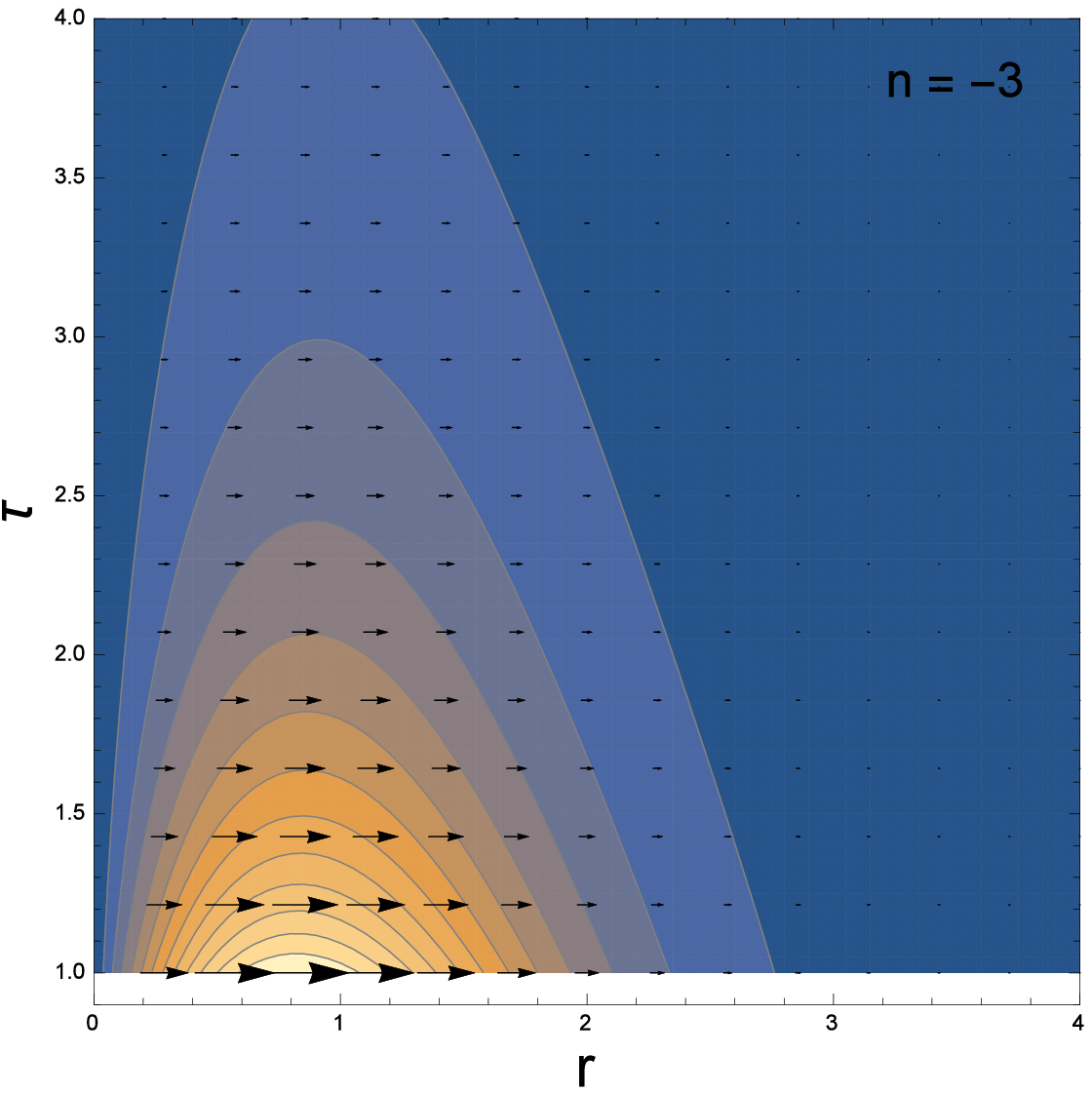}
        \includegraphics[width=0.6in]{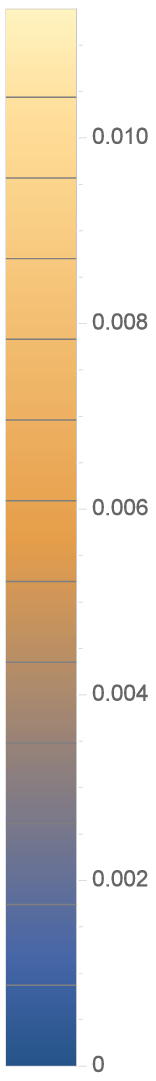} \\
		\end{array}$
	\end{center}
	\caption{Two-dimensional transverse fluid velocity  $v_r(\tau,r)$ is plotted with $n=-3$. The background colors represent the magnitude of $v_r$.}
\end{figure}

In Fig.~6, and Fig.~7, we make further comparisons for $v_r(\tau,r)$  at either fixed $r$ or fixed $\tau$ for the cases $n= -1$ and $n=-3$. They show  $\mid v_r(\tau,r) \mid$  becomes smaller in late times specially For $n=-3$ because of  faster decay of the magnetic field.  

\begin{figure}[h]
	\begin{center}$
		\begin{array}{cc}
		\includegraphics[width=3in]{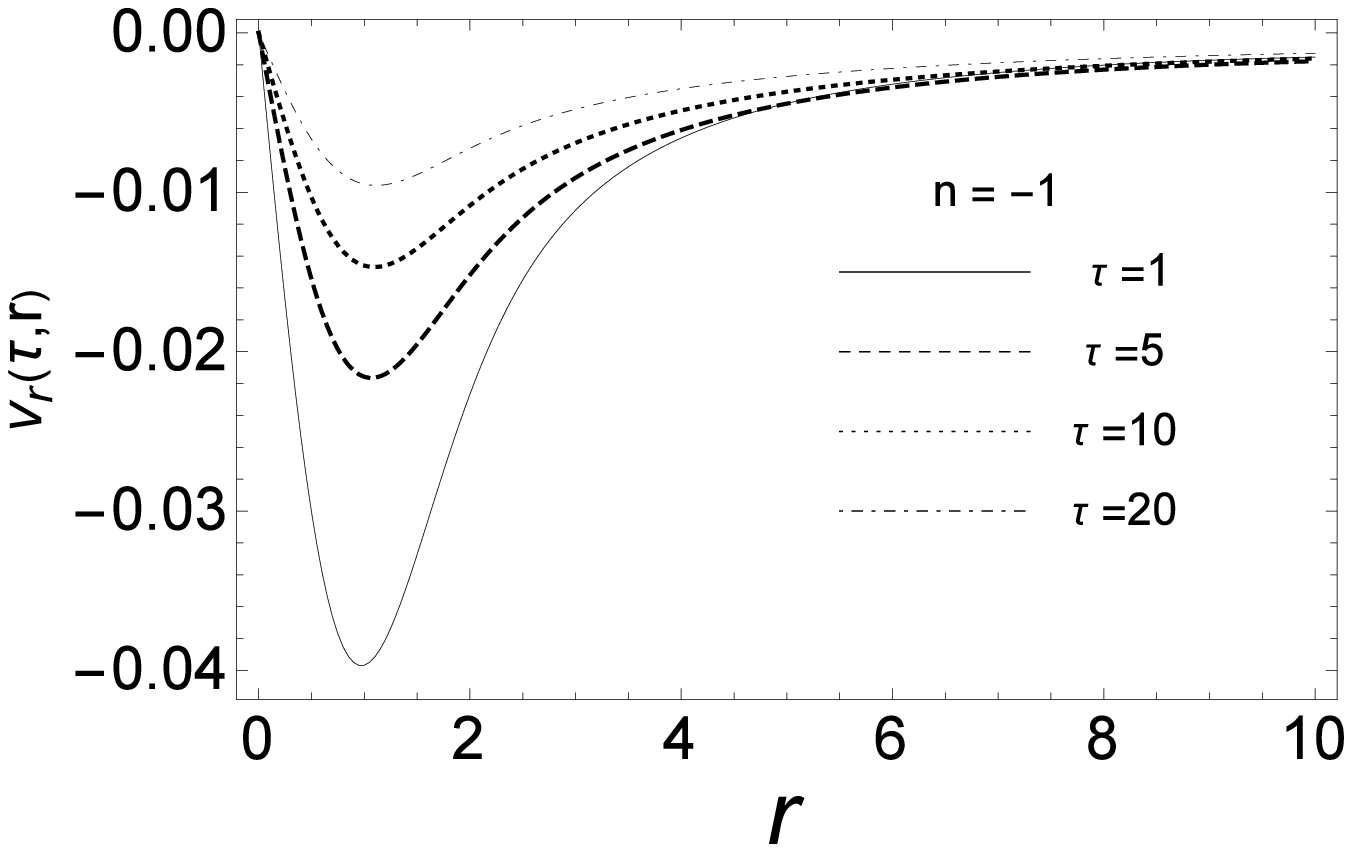}
        \includegraphics[width=3in]{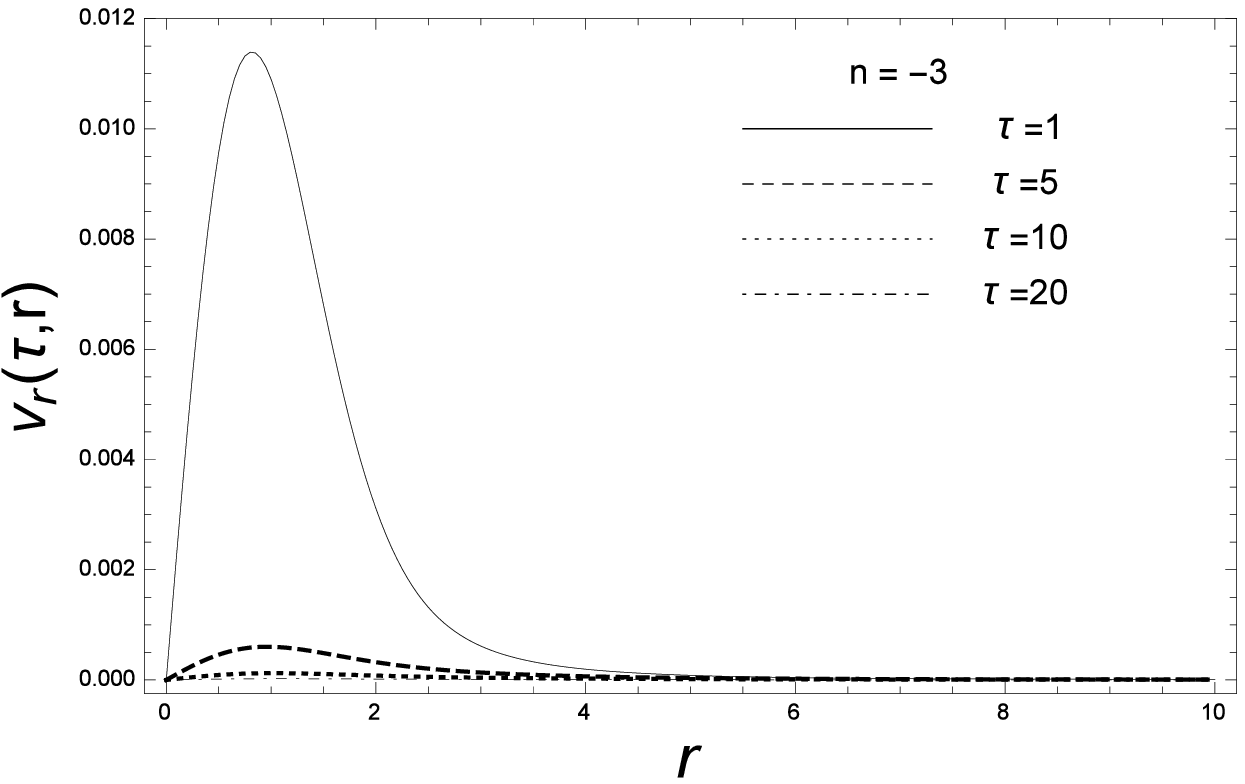}\\
		\end{array}$
	\end{center}
	\caption{ The transverse fluid velocity $v_r(\tau,r)$ in terms of cylindrical radial coordinate $r$  is plotted at different values of  proper time  $\tau$. Left panel is for $n=-1$  and right panel is for $n=-3$.}
\end{figure}

\begin{figure}[h]
	\begin{center}$
		\begin{array}{cc}
		\includegraphics[width=3in]{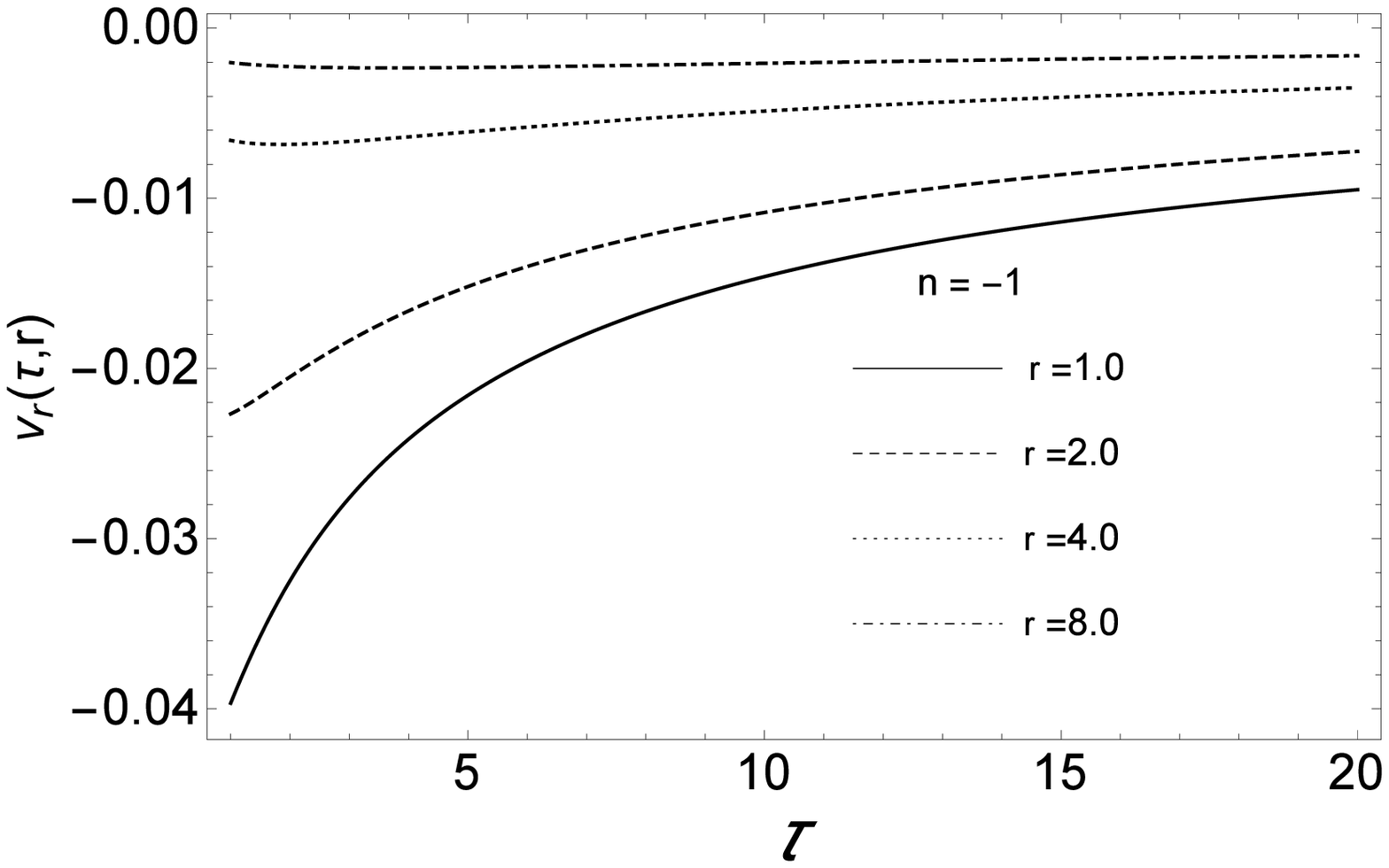} 
        \includegraphics[width=3in]{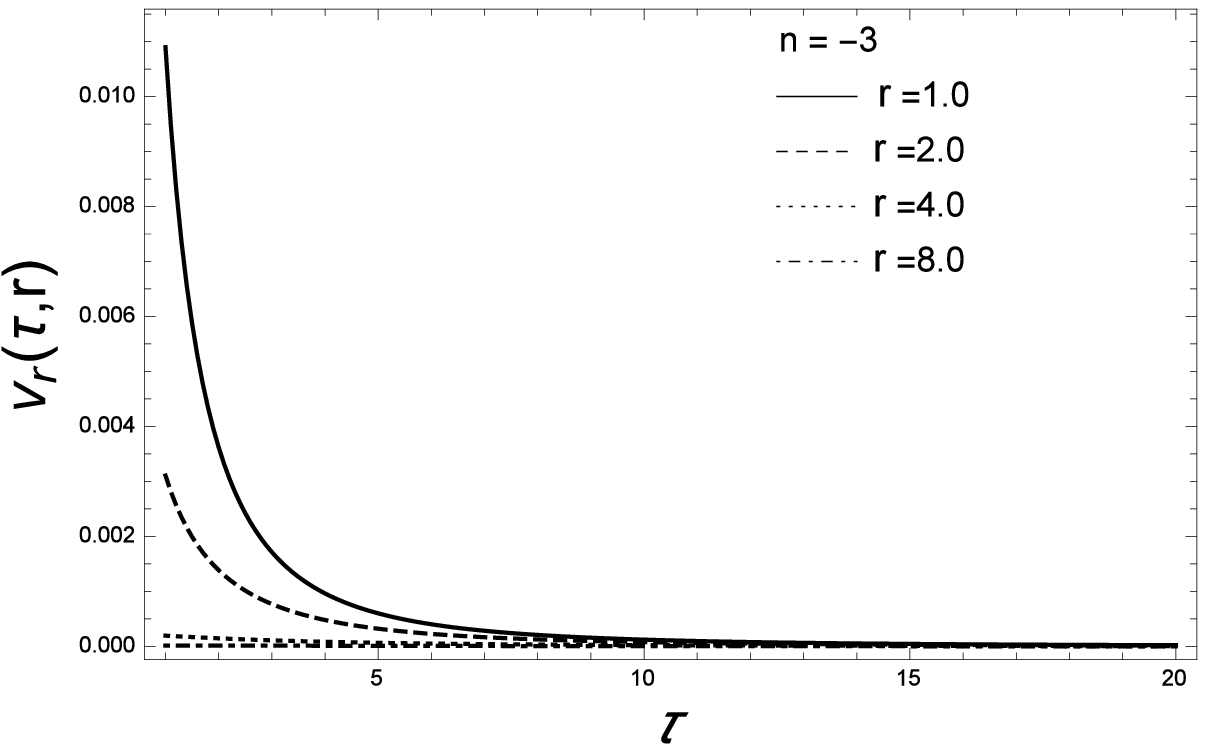} \\
		\end{array}$
	\end{center}
	\caption{The transverse fluid velocity $v_r(\tau,r)$ is  plotted  as function of  $\tau$  at different values of  radial coordinate $r$. Left panel is for $n=-1$  and right panel is for $n=-3$.}
\end{figure}

Figs.~8 and~9  indicate  the correction of energy density $\epsilon_1 (\tau,r)$ in terms  of  the radial coordinate  $r$ for different values of $\tau$ and in terms of  $\tau$ for different magnitudes  of  $r$ in the case $n=-1$ and $n=-3$, respectively. We remind the reader that the total energy density is $\epsilon=\epsilon_0+\epsilon_1 (\tau,r)$ and the second term is truly affected by the magnetic field. From Figs.~8 and ~9 For $n=-1$,  we deduce that the medium system’s element flows into the medium’s core, and the energy density decreases. In fact, the medium loses its energy in the form of magnetic field radiation. However, for $n=-3$, the magnetic field can give the additional contribution of energy to the medium to expand it in the transverse plane. 

\begin{figure}[h]
	\begin{center}$
\begin{array}{cc}
		\includegraphics[width=5in]{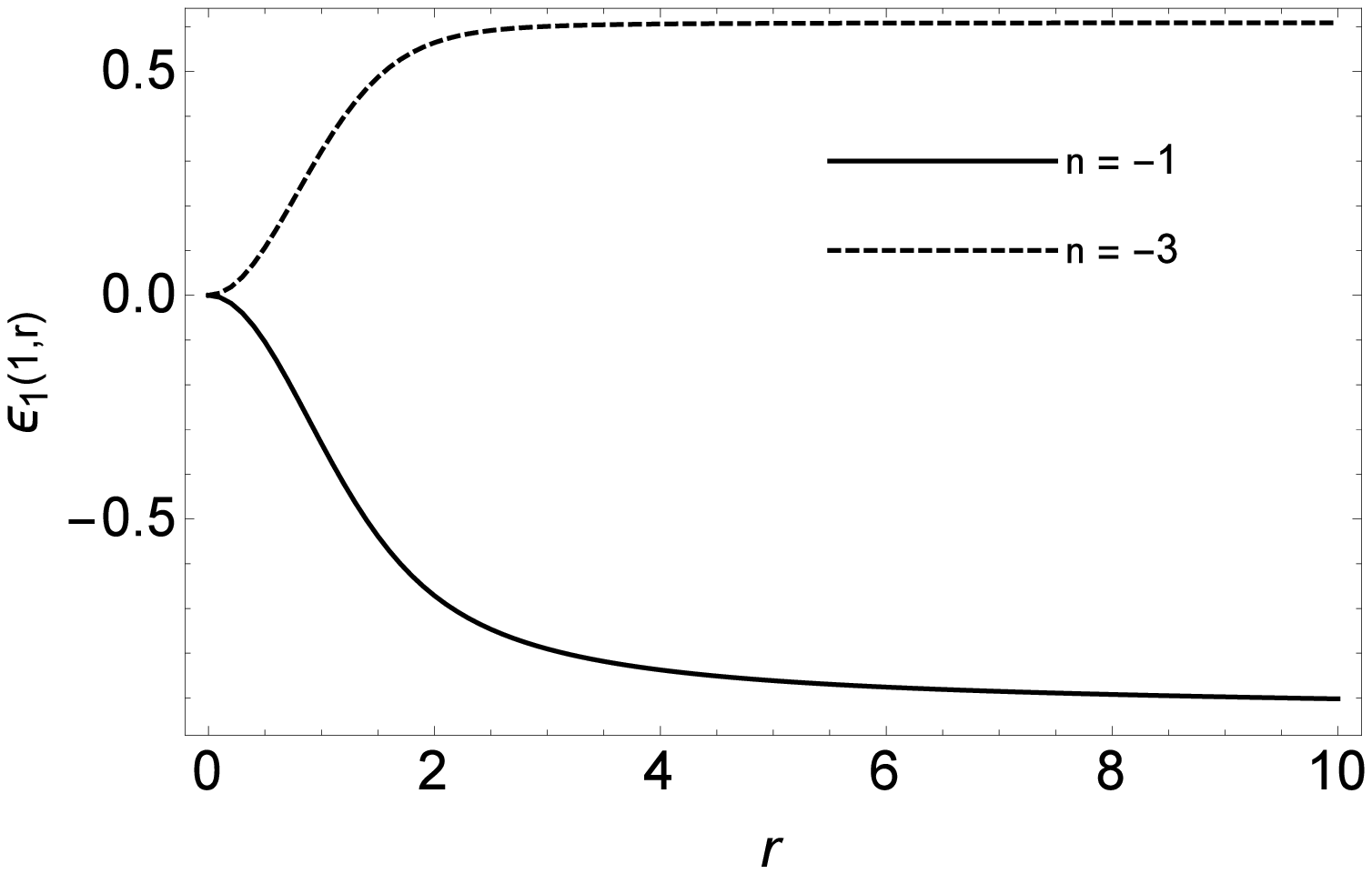}\\
\end{array}$
	\end{center}
	\caption{ The correction of energy density $\epsilon_1(\tau, r)$ ($GeV/fm^3$)  in terms of cylindrical radial coordinate  $r$  is plotted for two different values of $n=-1,-3$ at $\tau=1$ .}
\end{figure}

\begin{figure}[h]
	  \begin{center}$
      \begin{array}{cc}
      \includegraphics[width=5in]{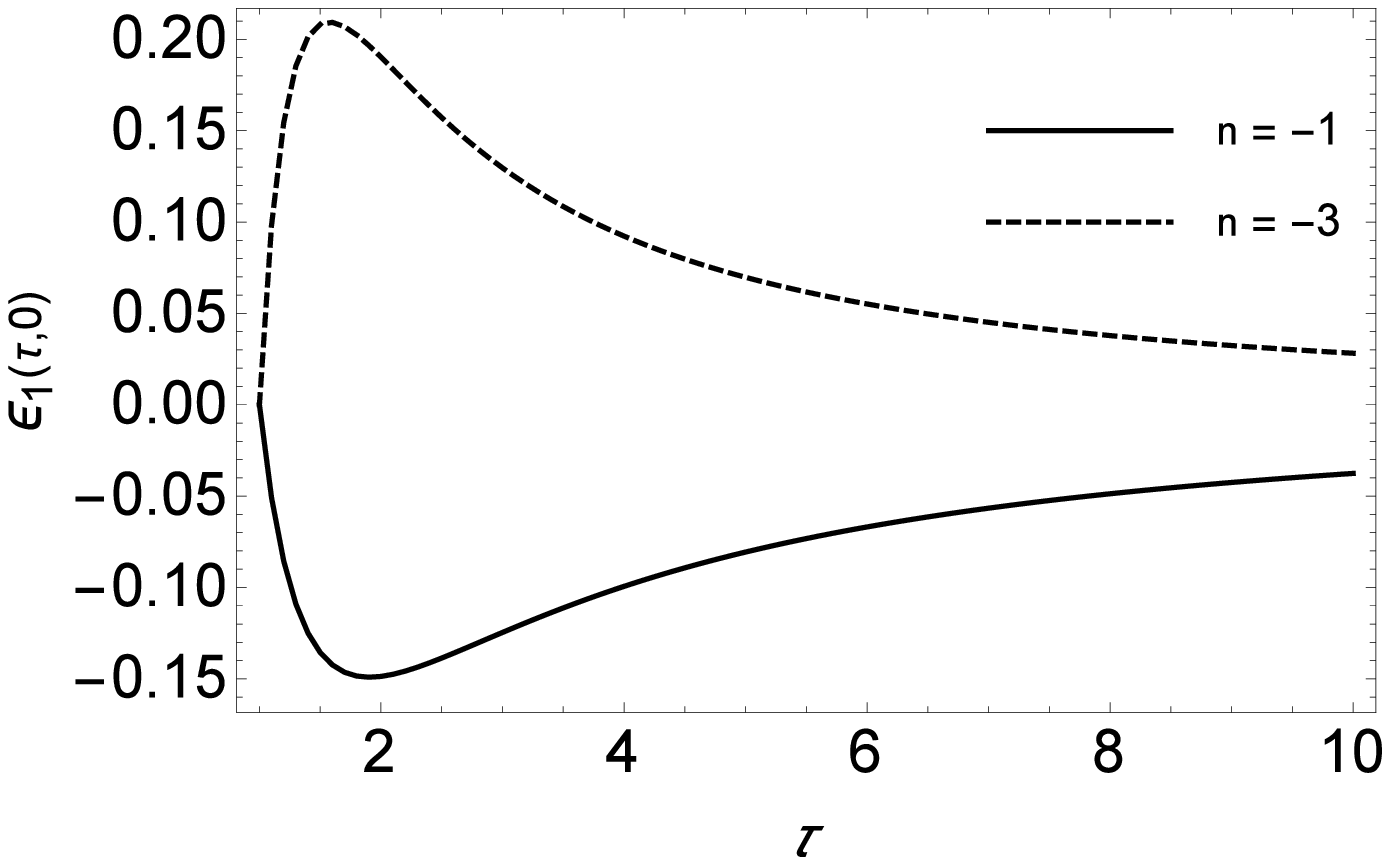} \\
      \end{array}$
	 \end{center}
	 \caption{The correction of energy density $\epsilon_1(\tau, r)$ ($GeV/fm^3$)  is  plotted  as function of  $\tau$  for two different values of $n=-1,-3$ at $r=0$.}
\end{figure}

In  Fig.~10, and Fig.~11 we show  further comparisons for  the correction of energy density $\epsilon_1 (\tau,r)$ in terms  of  radial coordinate  $r$ for different values of $\tau$ and in terms of   $\tau$ for different magnitudes  of  $r$ for  cases $n=-1$  and $n=-3$. We see $\epsilon_1 (\tau,r)$ is almost negative  for the case $n=-1$, and positive for the case $n=-3$.

In Fig.~12, and Fig.~13, we make further comparisons for $v_r(\tau,r)$  at either fixed $r$ or fixed $\tau$ for the cases $n<-2$ and $n>-2$. As shown in Fig.~12, when $\mid n\mid$ increases, $\mid v_r \mid$ at fixed $r$ becomes smaller in any times. Furthermore, as shown in Fig.~12 at $\tau= 1$, we find that the $\mid v_r \mid$ first increase from $r = 0$ and then turn over at intermediate $r$ and gradually decrease with $r$. For $n>-2$, the velocity profile has a similar shape compared with the cases for $n<-2$, while the direction becomes negative.  Considering  cases $n>-2$,  the fireball system   contracts in the   transverse plane, but for  $n<-2$ the medium expands in  the   transverse plane. Moreover, for the case $n=-2$  which is corresponded to  the ideal-MHD  the medium is stable in  the   transverse plane  
($v_r=0$).  We see a change of  the direction of the transverse velocity for $n<-2$
  or $n>-2$.  The transverse flow which  is created  by a external  magnetic field points inward for $n>-2$ and outward for$n<-2$ ,respectively. In the case $n=-2$
the transverse flow  created by the magnetic field is vanished. It is interesting if one considers $n$ depending on the properties of the fluid specially the electric conductivity ($\sigma$). Thus, we could consider the case $n=-2$ a perfect fluid with $\sigma$ goes to infinity. In this case, the decay of the magnetic field is correspond to $\frac{1}{\tau}$. Besides, the evolution of energy density and magnetic field are decoupled. In the case of $ n<-2$, one could  assume the fluid has a finite electric conductivity, and as $\mid n \mid$ increases  $\sigma$ goes to zero. Besides, the magnetic field decays faster then the expansion of medium.
 In the case of $ n>-2$, one could assume there is nonphysical fluid or  the electric conductivity is  negative. In this case the fluid contracts in the transverse plane, and  the magnetic field decays much slower than the expansion of fluid. We could  assume  there is  a static magnetic field  in a medium expanding along the $z$ direction.            
	
Figs~14, and~15 shows the correction energy density in terms of $\tau$ or $r$ at either fixed $r$ or fixed $\tau$ for the cases $n<-2$ and $n>-2$.
According to Fig.~14 and Fig.~15, the correction energy density starting from zero at proper time $\tau = 1 $ and for $n>-2$ the correction energy density is negative at any $r$ and $\tau$, while for $n<-2$ the correction energy density is positive at any $r$ and $\tau$. We could interpret above results as follow. For $n>-2$, the fluid  contracts, and medium  can give an additional contribution energy to the magnetic field. However, in the case $n<-2$
the magnetic field decays fast and the magnetic  energy is transferred to the fluid-element according to the energy-conservation law. Besides, the fluid expands radially. In special case $n=-2$  correspond to the situation of the ideal MHD limit in which the evolution of energy density and magnetic field are decoupled  which thus results in the absence of transverse  flow. In this case the magnetic field decays as $\frac{1}{\tau}$.

\subsection{Electromagnetic effect on the spectra }
In this subsection we investigate  effects of the magnetic field on the particle spectrum. We should notice our model is based on two  assumptions,
boost invariance along beam line, and rotation invariance in the transverse plane. Thus, we have  deduced  all  quantities  of interest only depend on the proper time and the radial coordinate $r$.  Therefore, we do expect our model fit with the experimental data only in mid rapidity and near central collisions.
In order to obtain the hadron spectra, we use the Cooper-Frye freeze out prescription  over the freeze-out surface~\cite{a35}:
\begin{eqnarray}
S=E\frac{d^3N}{dp^3}=\frac{dN}{p_Tdp_T dyd\varphi}=\int d\Sigma_\mu p^\mu \exp(\frac{-p^\mu u_\mu}{T_f})
\end{eqnarray}

Where  $T_f$ is the temperature at the freeze out surface, and $u^{\mu}$ is  the 4-velocity of the fluid.  The $d\Sigma_\mu$ is the element area  on the  isothermal freeze out surface in space-time.  The freeze out surface is  where the temperature of fluid is related to the energy density as $T\propto \epsilon^{1/4}$. It must satisfy $T(\tau, r)=T_f$.
In our convention, the area element perpendicular to the freeze out surface is given by:
 \begin{eqnarray}
 d\Sigma_\mu&=&(-1, R_f, 0, 0)\tau_f r dr d\varphi  d\eta.
 \end{eqnarray}
 The integration measure in the Cooper-Frye formula is:

 \begin{eqnarray}
 d\Sigma_\mu p^\mu&=&[-m_T\cosh(Y-\eta)+p_T R_f\cos(\varphi_p-\varphi)]\tau_f  dr d\varphi d\eta.
 \end{eqnarray}
  Moreover, the scalar product $p^{\mu}u_{\mu}$ in the Cooper-Frye formula is
 \begin{eqnarray}
 p^\mu u_\mu&=&-m_T\cosh(Y-\eta)u_\tau+p_T\cos(\varphi_p-\varphi)u_r,
 \end{eqnarray}
 where $R_f\equiv -\frac{\partial \tau}{\partial r}=\frac{\partial_r T}{\partial_\tau T}\mid_{T_f}$.
 Here $\tau=\sqrt{t^2-z^2}$, $ r$, and  $\eta=\frac{1}{2}\log\frac{t+z}{t-z}$,  are   the longitudinal proper time,  the transverse
 (cylindrical) radius, and the longitudinal rapidity
 (hyperbolic arc angle),  respectively. Similarly $u_r$ is the transverse flow velocity and $\varphi$ is its asimuthal angle.  $\varphi_p$ is the azimuthal angle in momentum space. Besides,   $p_T$, $m_T=\sqrt{m^2+p_T^2}$, and  $Y$ are the detected transverse momentum,   the corresponding
 transverse mass, and the observed longitudinal rapidity respectively. Thus, the final expression for the CF formula is:
 \begin{eqnarray}
 S=\frac{g_i}{2\pi^2}\int_0^{x_f}\  r\ \tau_f(r)\ dr\
 \Big[m_T K_1(\frac{m_Tu_\tau}{T_f})I_0(\frac{m_Tu_r}{T_f})+p_T R_f
 K_0(\frac{m_Tu_\tau}{T_f})I_1(\frac{m_Tu_r}{T_f})\Big]
 \label{spectrum}
 \end{eqnarray}
 Where $\tau_f(r)$  and  $g_i$  are the solution of the $T(\tau_f, r)=T_f$ , and   the degeneracy factor for particles respectively.

  The above integral over $r$ on the freeze-out surface is evaluated numerically. Then,    the spectra of hadrons is obtained  as a function of
  $p_T$. The results for the charged pion  spectra
  are presented in Figs.~\ref{fig:figpion1} and \ref{fig:figpion}.

 Figs.~\ref{fig:figpion1} and \ref{fig:figpion} are demonstrated the hadron spectrum of pions.  We choose the $T = 130$ MeV isotherm as
 the freezeout surface in Fig ~\ref{fig:figpion1} for three different
 values of  the free parameter $r_0$. The  figure is the comparison of the
 spectrum of  pions (black, bottom)
  as a function of transverse momentum $p_T$  resulting from our hydrodynamic solution   with experimental results  for pions (top)  obtained at PHENIX ~\cite{a36}. We see the smallest value of $r_0$
 slightly brings the calculation closer to the experimental data. In Fig~
 \ref{fig:figpion1},  we choose three different values   of  the freeze out temperature ($140$, $150$ and $160$ MeV) and compared with experimental data  at PHENIX ~\cite{a36}
 in central collisions $0-5\%$.

 Our spectra appear to underestimate the experimental data, but their behavior with $p_T$ has the correct trend of
 a monotonically decrease. The highest value of the freeze out temperature we employed (as suggested, e.g. in Ref.~\cite{a37})
 slightly brings the calculation closer to the experimental data; however it also shows a kind of saturation phenomenon
 and points to the need of including other effects not considered in the present calculation.


\begin{figure}[h]
	\begin{center}$
		\begin{array}{cc}
		\includegraphics[width=3in]{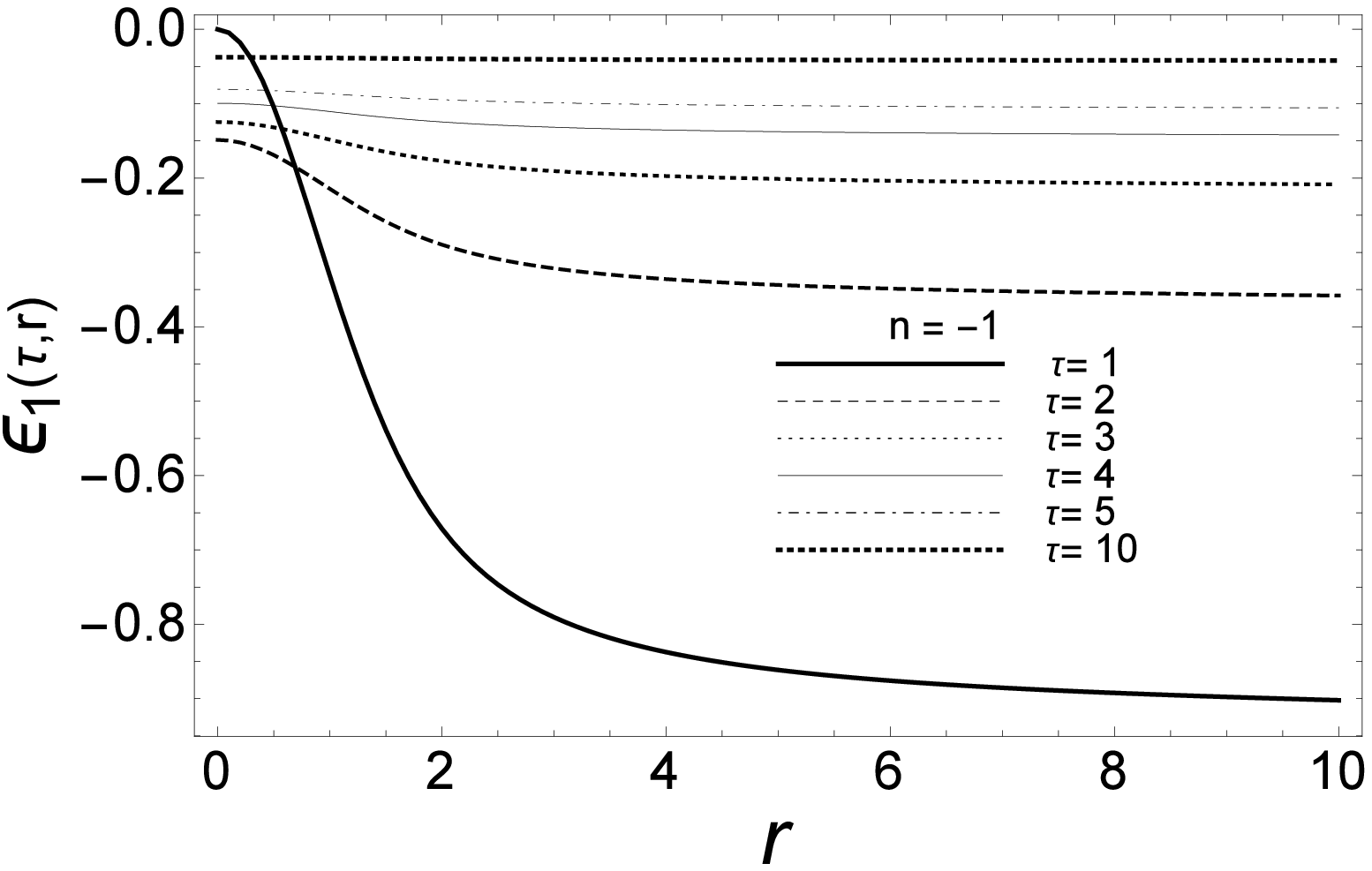} 
        \includegraphics[width=3in]{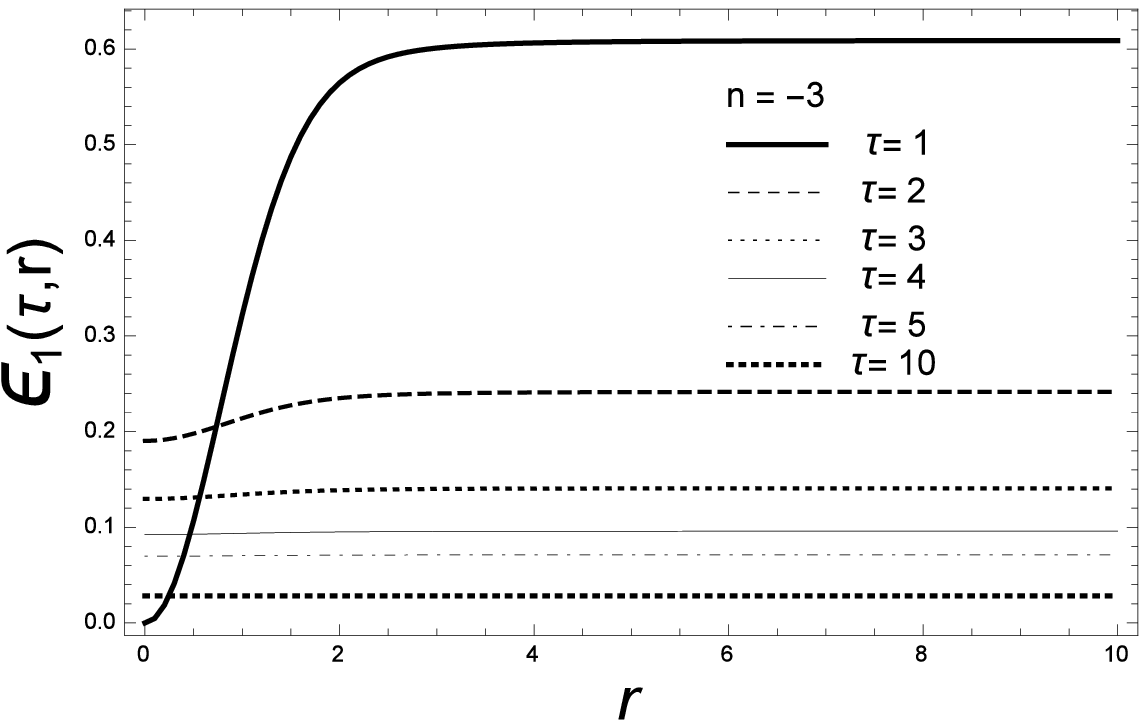}\\ 
		\end{array}$
	\end{center}
	\caption{The correction energy density  $\epsilon_1$ ($GeV/fm^3$) is plotted as a function of radial coordinate  $r$ for different values of $\tau$. Left panel $n=-1$  and right panel $n=-3$.}
\end{figure}

\begin{figure}[h]
	\begin{center}$
		\begin{array}{cc}
		\includegraphics[width=3in]{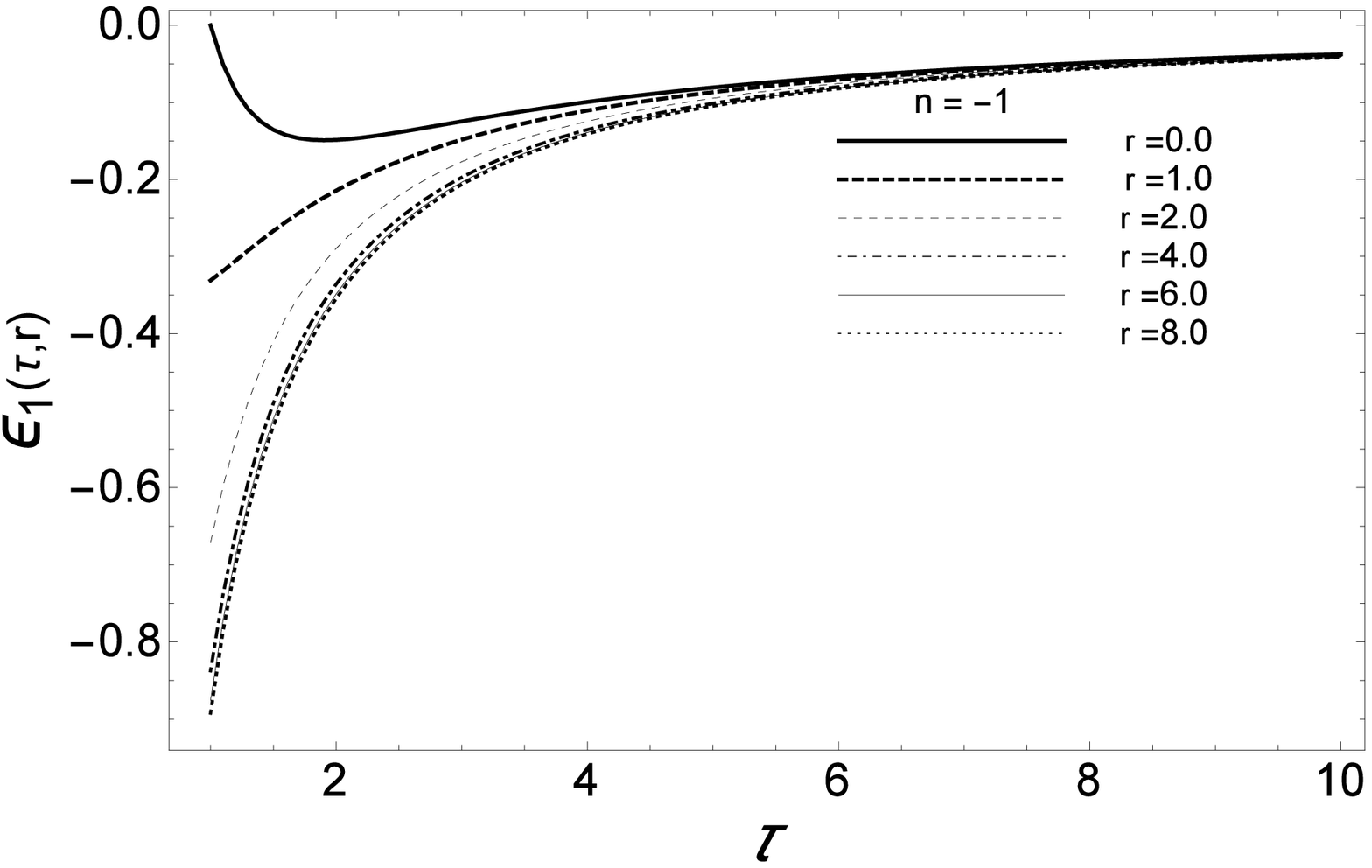}
        \includegraphics[width=3in]{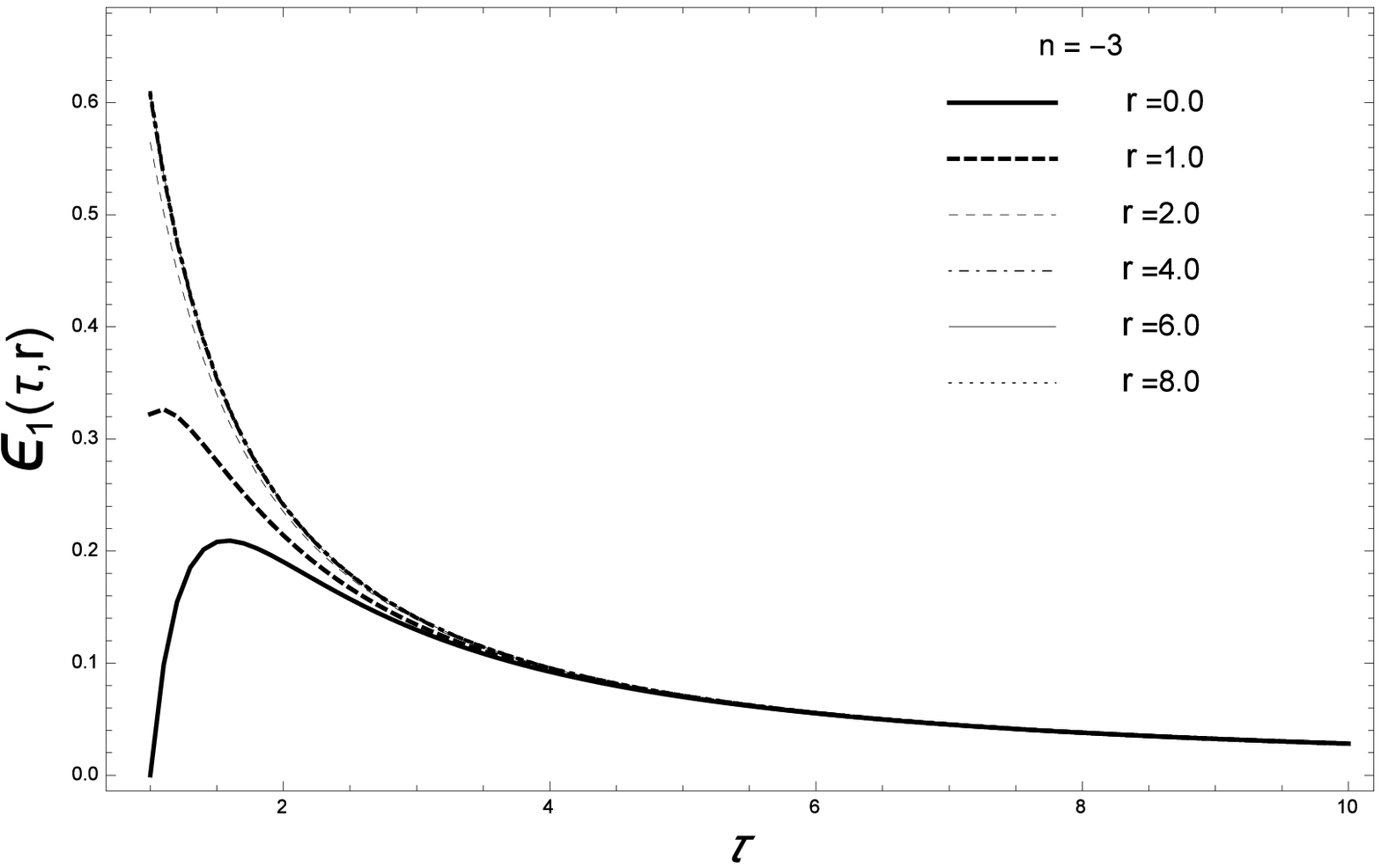}\\
		\end{array}$
	\end{center}
	\caption{The correction energy density    $\epsilon_1 $ ($GeV/fm^3$) as a function of proper time  $\tau$ for different values of radial coordinate  $r$. Left panel $n=-1$  and right panel $n=-3$.}
\end{figure}




\begin{figure}[h]
	\begin{center}$
		\begin{array}{cc}
		\includegraphics[width=5in]{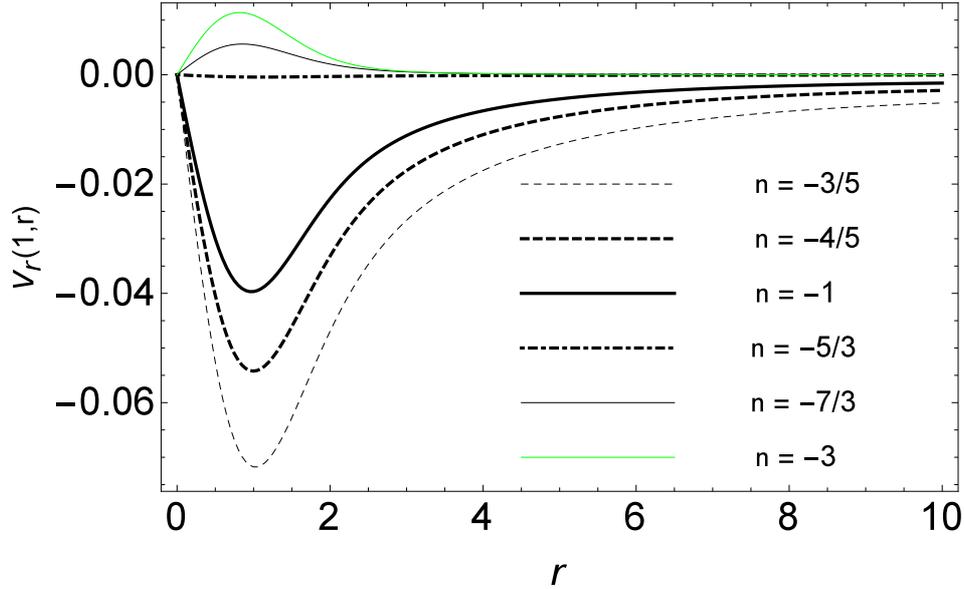}\\
		\end{array}$
	\end{center}
	\caption{The fluid velocity  $v_r(\tau, r)$ is plotted in terms of  $r$  at  $\tau = 1$ with different values of $n$.}
\end{figure}

\begin{figure}[h]
	\begin{center}$
		\begin{array}{cc}
		\includegraphics[width=5in]{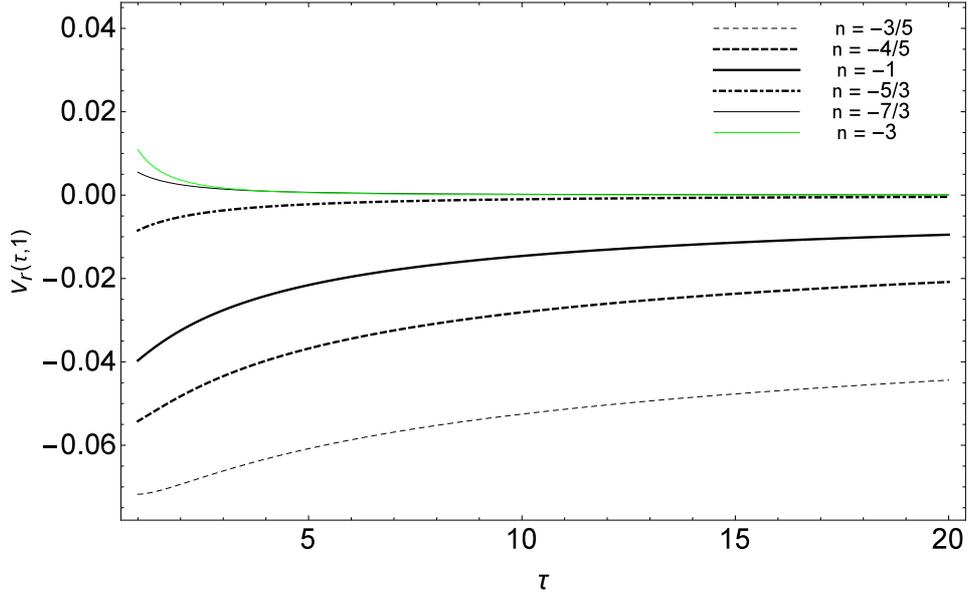} \\
		\end{array}$
	\end{center}
	\caption{The fluid velocity $v_r(\tau, r)$ in terms of  $\tau$ plot at $r = 1$ with different values of $n$.}
\end{figure}

\begin{figure}[h]
	\begin{center}$
		\begin{array}{cc}
		\includegraphics[width=5in]{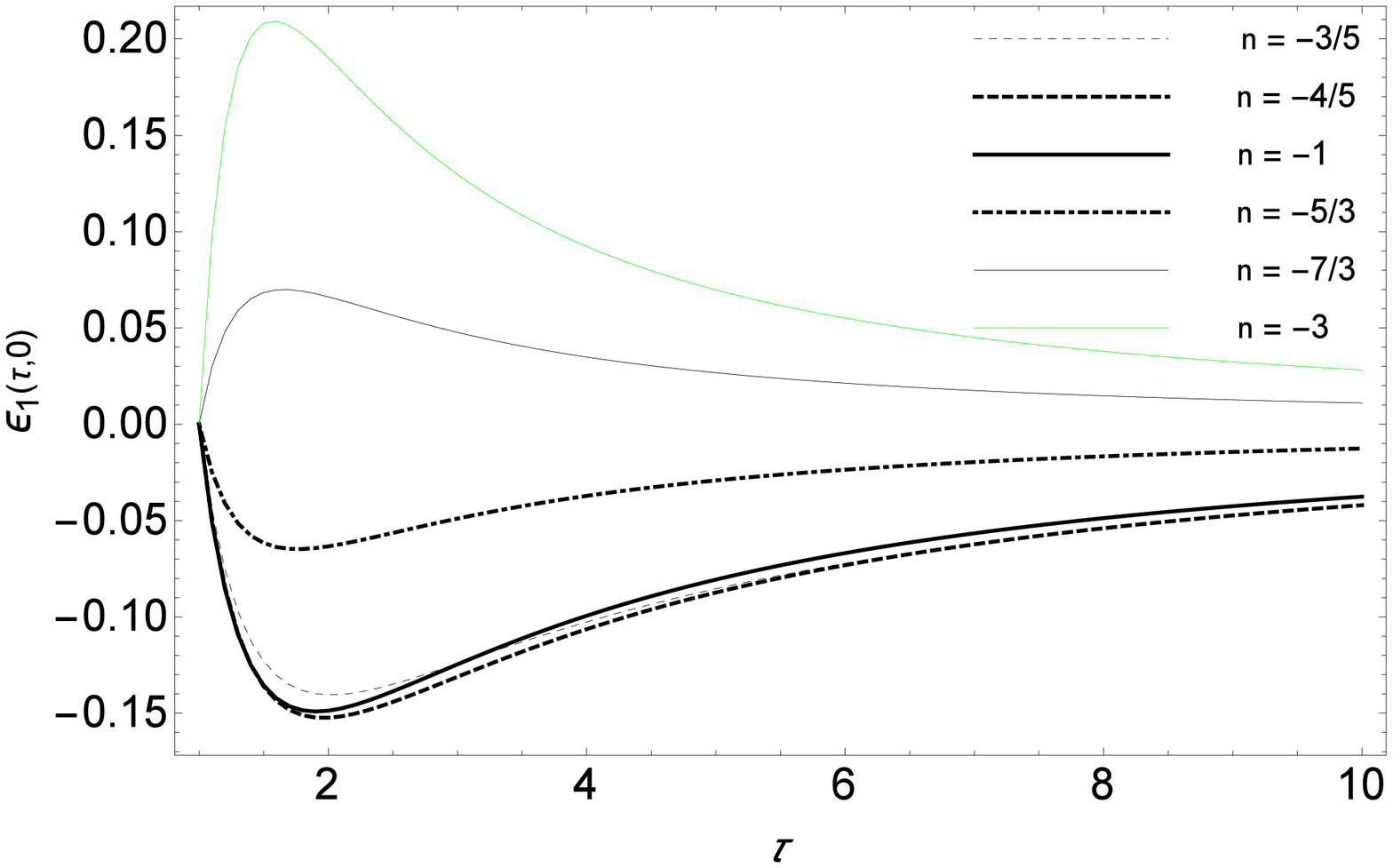} \\
		\end{array}$
	\end{center}
	\caption{ The correction energy density $\epsilon_1(\tau, r)$ ($GeV/fm^3$) is  plotted in terms of  $\tau$  at $r = 0$ with different values of $n$.}
\end{figure}

\begin{figure}[h]
	\begin{center}$
		\begin{array}{cc}
		\includegraphics[width=5in]{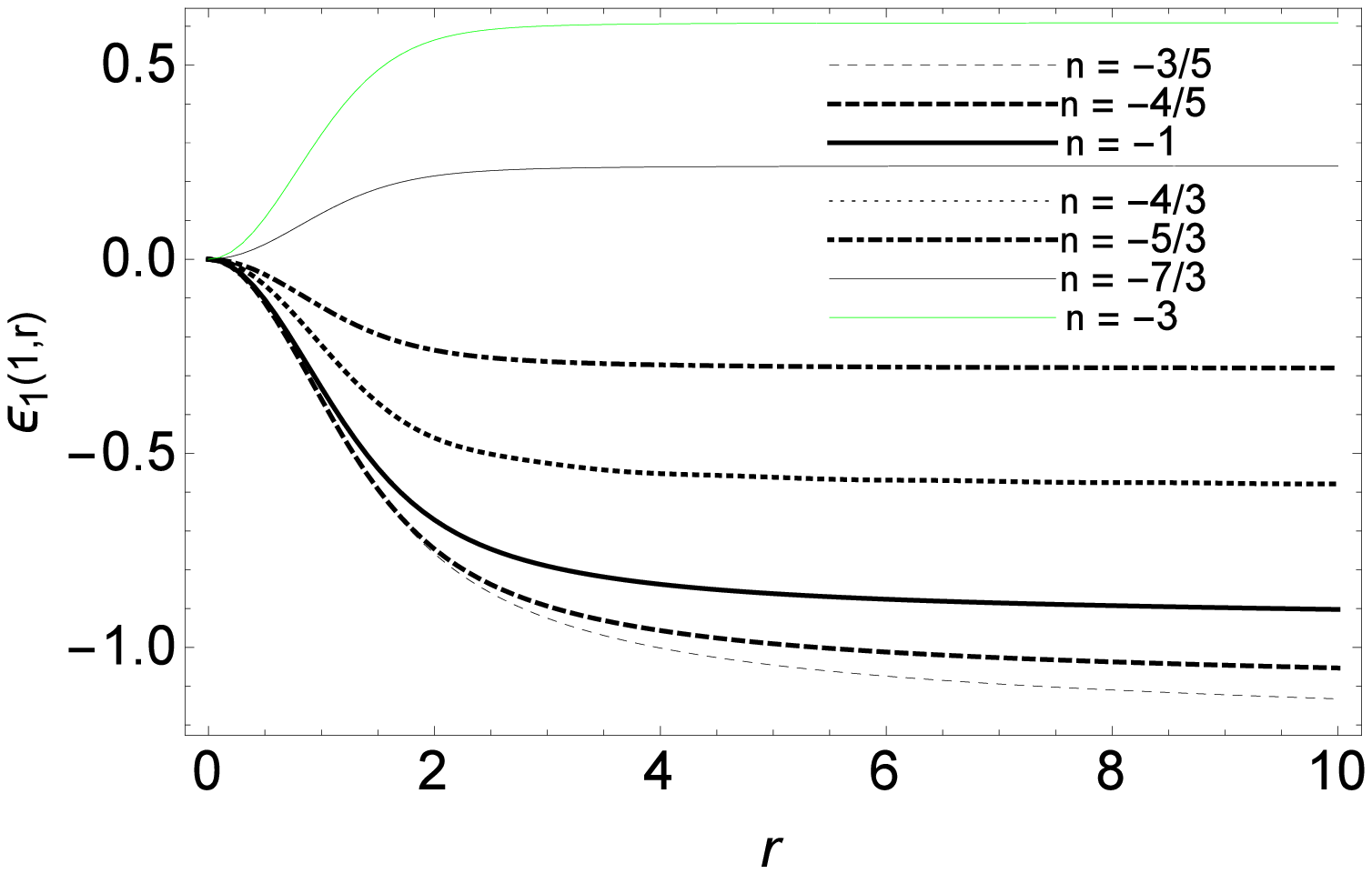} \\
		\end{array}$
	\end{center}
	\caption{The correction energy density $\epsilon_1(\tau, r)$ ($GeV/fm^3$)   is plotted in terms of  $r$  at $\tau = 1$ with different values of $n$.}
\end{figure}

\begin{figure}[h]
	\begin{center}
		\includegraphics[width=5in]{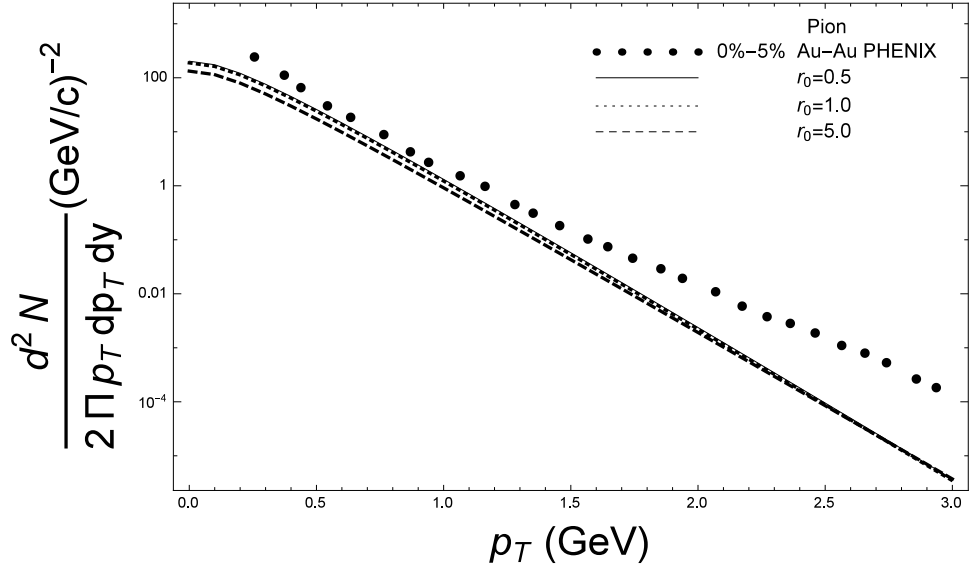} \\
	\end{center}
	\caption{Pion transverse spectrum from central Au-Au collisions   with value of $n=-1$. The black solid,  thin dashed, and thick dashed lines 	are  corresponded to $r_0=0.5, 1$ and $5$,
		respectively. Black dot line  is  PHENIX data.  }
	\label{fig:figpion1}
\end{figure}

\begin{figure}[h]
	\begin{center}
		\includegraphics[width=5in]{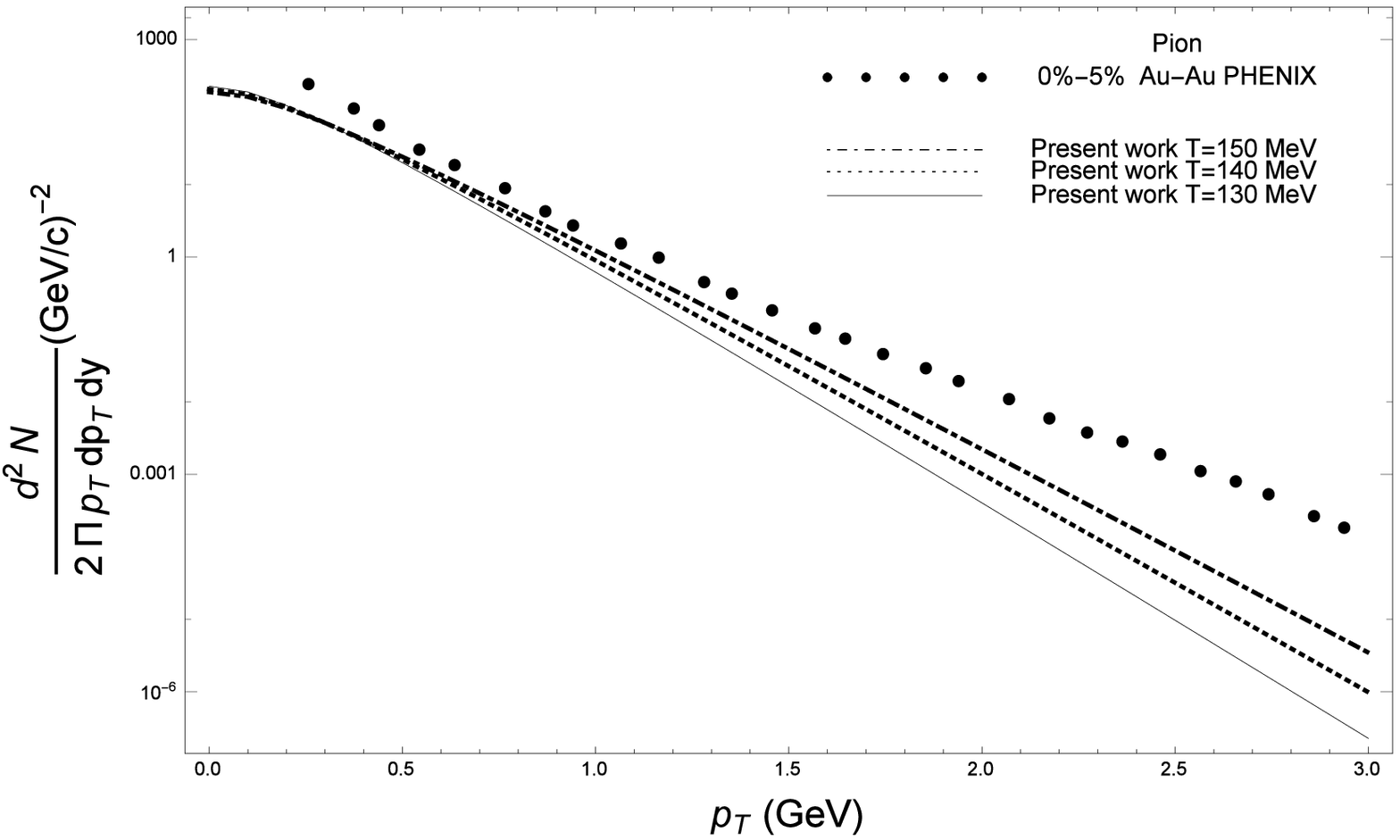} \\
	\end{center}
	\caption{Pion transverse spectrum from central Au-Au collisions with value of $n=-1$. The black  solid, dashed, and dash-dotted     lines are corresponded to a
		freeze out temperature of 140, 150 and 160 MeV, respectively. The black dots are  PHENIX data.}
	\label{fig:figpion}
\end{figure}


\section{Conclusions and outlook}
In the present work, we  have investigated  heavy ion collisions in the presence of a transverse external magnetic field  on the special case of a (1 + 2) dimensional longitudinally boost-invariant fluid expansion. We have solved  magneto-hydrodynamic equations in presence of a external magnetic field perturbatively.  We work in  Milne coordinates, and  the medium is assumed  boost-invariant along the $z$ direction. We remark that in  Ref. \cite{a28} it been supposed that the external magnetic field is located in the transverse plane as $b_\mu= (0, 0, b_\varphi, 0)$ where $b_\mu b^{\mu}= b^2$ is defined. Their study was in a simple setup, which includes an azimuthal magnetic field in the matter distribution. In current work we modify the framework, and assume the external magnetic field has  two components in transverse plane. Working in cylinder coordinates in transverse plane, we have supposed that $b_\mu= (0, b_r, b_\varphi, 0)$; however, we have  assumed that square of the magnetic field only depends on $r$. This extra assumption simplifies the energy conservation and Euler equations. We  have shown that by combining  magnetic field with the boost symmetry along the beam direction, a radial flow perpendicular to the beam axis is created and the energy density of the fluid is altered. Although we have chosen the Gaussian distribution as one particular example for the space-dependent magnetic field, the same approach can be applied to other spatial distribution which can be approximated by a series of Bessel functions. Since the analytic expressions of each moment is found, one can directly compute the transverse velocity and correction on energy density by just inputting the Bessel coefficients of series. The energy conservation and Euler equations reduced to two coupled differential equations, which could be solved analytically in the weak-field approximation. We have demonstrated  in detail how the fluid velocity and energy density are modified by the magnetic field. For the solutions obtained by our numerical calculations we have assumed  an initial energy density of the fluid at time $\tau = 1$ fixed to  $B^2/\epsilon_c=0.6$. We have  considered  different decays with time of the magnetic field: $\tau^n/2$, with $n<0$. A visual presentation of the flow for $n =-1$ and $n=-3$  can be find in Figs.~2 and 3 and for different values of $n$ in Figs.~12 and 13.\\
We stress that the present work presents an approximated calculation which can be useful for cross checking current and future numerical calculations in some limiting region. Indeed, the effect of such a scenario on hadronic flow in heavy ion collisions requires more pragmatic debates. Moreover our perturbative solutions could be used as initial conditions in the late time for  future  numerical calculations.

\end{document}